\documentclass[aps,pra,twocolumn,superscriptaddress]{revtex4-2}

\usepackage{graphicx}
\usepackage{epstopdf}
\usepackage{amsmath}
\usepackage{amssymb}
\usepackage{amsfonts}
\usepackage{mathrsfs}
\usepackage{theorem}
\usepackage{bm}
\usepackage{url}
\usepackage[T1]{fontenc}
\usepackage{csquotes}
\MakeOuterQuote{"}
\usepackage{multirow}
\usepackage{makecell}
\usepackage{array}
\usepackage{booktabs}

\usepackage{algorithm}
\usepackage{algorithmicx}
\usepackage{algpseudocode}

\usepackage{dcolumn}
\usepackage{color}
\usepackage{xcolor}

\usepackage[colorlinks]{hyperref}
\hypersetup{
	colorlinks = true,
	urlcolor = {blue},
	citecolor = {blue},
	linkcolor= {blue}
}

\def\vec#1{\bm{#1}}

\newcommand{\Tr}{\mathrm{Tr}}

\newcommand{\dd}{\mathrm{d}}

\newcommand{\id}{1\!\!1}

\newcommand{\ideal}{\mathrm{ideal}}
\newcommand{\ex}{\mathrm{exp}}

\newcommand{\rmi}{\mathrm{i}}
\newcommand{\rme}{\mathrm{e}}

\newcommand{\proj}{\mathrm{proj}}
\newcommand{\opt}{\mathrm{opt}}

\def\<{\langle}  %% overiding the original command \<
\def\>{\rangle}  %% overiding the original command \>
\newcommand{\ket}[1]{| #1\>}
\newcommand{\bra}[1]{\< #1|}

\newcommand{\braket}[2]{\langle #1  |#2\rangle}
\newcommand{\ketbra}[2]{|#1\rangle\langle #2|  }

\def\eqref#1{\textup{(\ref{#1})}}  %% overiding the original command \eqref
\newcommand{\eref}[1]{Eq.~\textup{(\ref{#1})}}

\newcommand{\fref}[1]{Fig.~\ref{#1}}
\newcommand{\Fref}[1]{Figure~\ref{#1}}

\newcommand{\rcite}[1]{Ref.~\cite{#1}}

\newcommand{\up}[1]{\mathrm{up}}

\usepackage{mathtools,mleftright}
\mleftright
\begin{document}
\title{A single programmable photonic circuit for universal quantum measurements}
\date{\today}

\author{Wen-Zhe Yan} 
\thanks{These authors contributed equally to this work.}
\affiliation{Laboratory of Quantum Information, University of Science and Technology of China, Hefei, 230026, China}
\affiliation{Anhui Province Key Laboratory of Quantum Network, University of Science and Technology of China, Hefei, 230026, China}
 
\author{Lan-Tian Feng} 
\thanks{These authors contributed equally to this work.}
\affiliation{Laboratory of Quantum Information, University of Science and Technology of China, Hefei, 230026, China}
\affiliation{Anhui Province Key Laboratory of Quantum Network, University of Science and Technology of China, Hefei, 230026, China}
\affiliation{CAS Center For Excellence in Quantum Information and Quantum Physics, University of Science and Technology of China, Hefei, 230026,  China}
\affiliation{Hefei National Laboratory, Hefei, 230088, China}
    
\author{Zhibo Hou} 
\thanks{These authors contributed equally to this work.}
\email{houzhibo@ustc.edu.cn}
\affiliation{Laboratory of Quantum Information, University of Science and Technology of China, Hefei, 230026, China}
\affiliation{Anhui Province Key Laboratory of Quantum Network, University of Science and Technology of China, Hefei, 230026, China}
\affiliation{CAS Center For Excellence in Quantum Information and Quantum Physics, University of Science and Technology of China, Hefei, 230026,  China}
\affiliation{Hefei National Laboratory, Hefei, 230088, China}

\author{Yuan-Yuan Zhao} 
\thanks{These authors contributed equally to this work.}
\affiliation{Quantum Science Center of Guangdong-Hong Kong-Macao Greater Bay Area, Shenzhen 518045, China}

\author{Carles Roch i Carceller} 
\affiliation{Physics Department and NanoLund, Lund University, Box 118, 22100 Lund, Sweden}

\author{Armin Tavakoli} 
\email{armin.tavakoli@fysik.lu.se}
\affiliation{Physics Department and NanoLund, Lund University, Box 118, 22100 Lund, Sweden}

\author{Huangjun Zhu}
\email{zhuhuangjun@fudan.edu.cn}
\affiliation{State Key Laboratory of Surface Physics, Department of Physics, and Center for Field Theory and Particle Physics, Fudan University, Shanghai 200433, China}
\affiliation{Institute for Nanoelectronic Devices and Quantum Computing, Fudan University, Shanghai 200433, China}
\affiliation{Shanghai Research Center for Quantum Sciences, Shanghai 201315, China}
\affiliation{Hefei National Laboratory, Hefei, 230088, China}

\author{Guang-Can Guo}
\affiliation{Laboratory of Quantum Information, University of Science and Technology of China, Hefei, 230026, China}
\affiliation{Anhui Province Key Laboratory of Quantum Network, University of Science and Technology of China, Hefei, 230026, China}
\affiliation{CAS Center For Excellence in Quantum Information and Quantum Physics, University of Science and Technology of China, Hefei, 230026,  China}
\affiliation{Hefei National Laboratory, Hefei, 230088, China}

\author{Xi-Feng Ren}
\email{renxf@ustc.edu.cn}
\affiliation{Laboratory of Quantum Information, University of Science and Technology of China, Hefei, 230026, China}
\affiliation{Anhui Province Key Laboratory of Quantum Network, University of Science and Technology of China, Hefei, 230026, China}
\affiliation{CAS Center For Excellence in Quantum Information and Quantum Physics, University of Science and Technology of China, Hefei, 230026,  China}
\affiliation{Hefei National Laboratory, Hefei, 230088, China}

\author{Guo-Yong Xiang}
\email{gyxiang@ustc.edu.cn}
\affiliation{Laboratory of Quantum Information, University of Science and Technology of China, Hefei, 230026, China}
\affiliation{Anhui Province Key Laboratory of Quantum Network, University of Science and Technology of China, Hefei, 230026, China}
\affiliation{CAS Center For Excellence in Quantum Information and Quantum Physics, University of Science and Technology of China, Hefei, 230026,  China}
\affiliation{Hefei National Laboratory, Hefei, 230088, China}
 
\begin{abstract}
    \noindent\textbf{Abstract}\\
    Programmable photonic quantum processors face a critical challenge: despite significant advances in quantum state preparation and manipulation, measurements remain limited to projective techniques. Here, we demonstrate a programmable measurement processor that overcomes this limitation by enabling arbitrary quantum measurements within a scalable circuit framework. Our large-scale integrated photonic architecture achieves precise coherent control of ancillary quantum systems, realizing a universal four-dimensional quantum measurement device. We benchmark the processor by performing measurement tomography on 100 randomly selected measurements, achieving an average fidelity of 97.7\%. The processor's performance exceeds the theoretical limits of projective measurements in three key quantum information tasks: state discrimination (with 23 times lower error), state estimation (with 10.6\% higher fidelity), and randomness generation (with 37\% more randomness yield), demonstrating its high operational quality. This work establishes a fully programmable quantum measurement processor, advancing the development of universal quantum operations for photonic quantum information processing by providing the key missing component.
	
\end{abstract}
\maketitle

\noindent\textbf{Introduction}\\
\noindent The evolution from specialized quantum experiments to programmable quantum architectures represents a pivotal {advancement} in quantum technologies. In integrated photonics \cite{wang20integrated,Feng22Silicon},  this ambition has driven remarkable progress toward universal control, i.e.,~the ability to generate and manipulate quantum states of light at will via reconfigurable circuit interferometry \cite{wang2018multidimensional, carolan2015universal, bogaerts2020programmable}. However,  not all modules required for a photonic quantum processing unit have advanced equally. Measurement devices have remained more limited, with full programmability  having been achieved only for standard basis measurements, also known as projective measurements \cite{chi2022programmable}.

\begin{figure*}[tpb]
	\centering	
	\includegraphics[width=\linewidth]{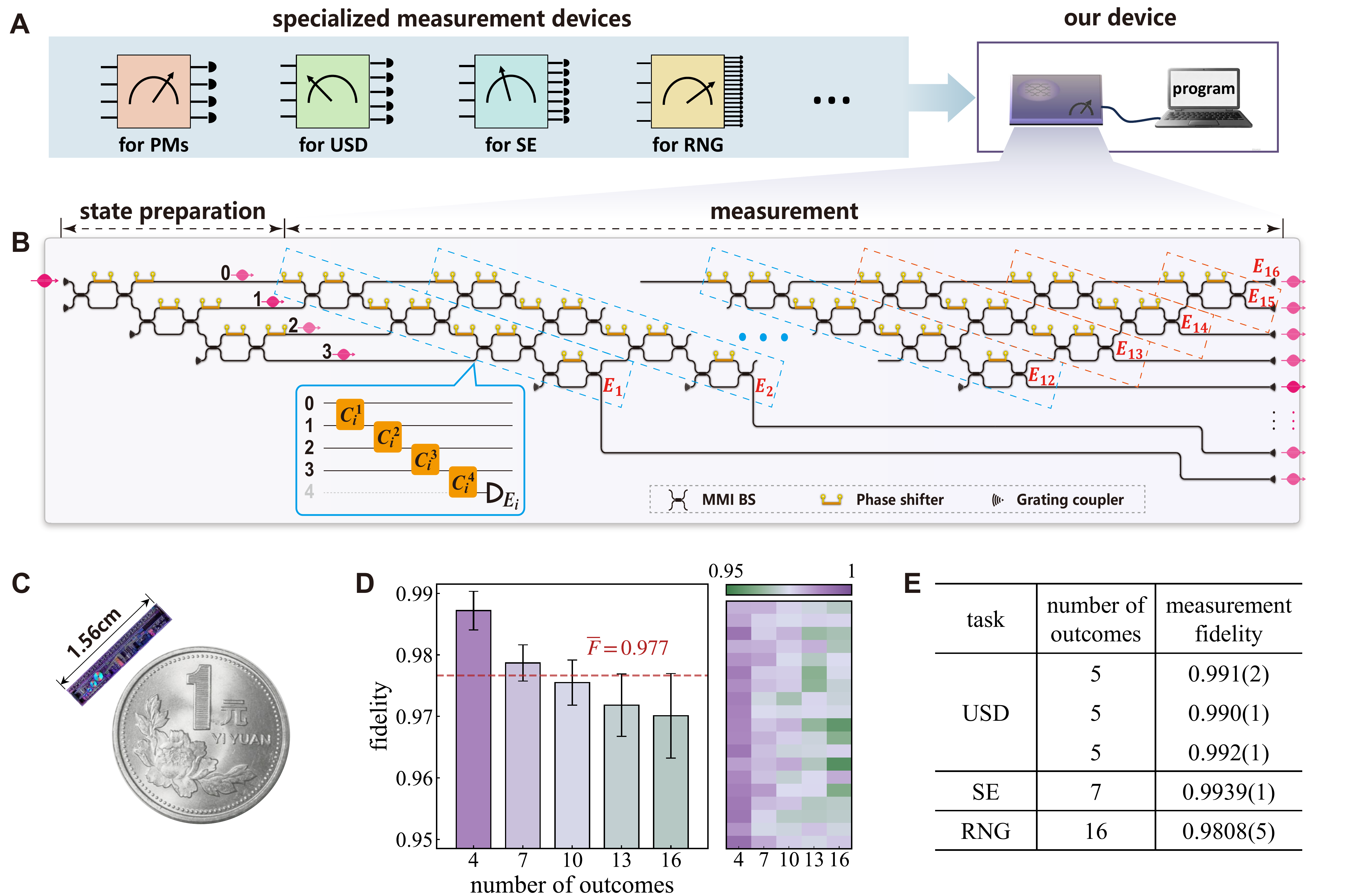}
	\caption{\label{fig: setup} \textbf{Universal quantum measurement circuit and its characterization.}
    (\textbf{A})~Conceptual illustration.
    Specialized devices for implementing different types of quantum measurements can be replaced with a single programmable device capable of implementing any measurement. 
    (\textbf{B})~Experimental setup. An input single photon is prepared in an arbitrary path-encoded four-dimensional state using three Mach-Zehnder interferometers (MZIs). The  measurement circuit can implement an arbitrary measurement on it through a cascade of 15 periodic modules.  Each module contains four MZIs tied to four two-dimensional unitary operators $C_i^j$ ($j=1,\dots,4$) and is configured to realize the outcome associated with the measurement operator $E_i$. The last three modules can be further simplified by removing the trivial MZIs.
    (\textbf{C})~Photograph of our chip alongside a coin.
    (\textbf{D})~Fidelities of 100 randomly sampled measurements determined by measurement tomography. Left: The average fidelity and its standard deviation for each set of 20 measurements with a specific number of outcomes. Right: all 100 individual fidelities.    
    (\textbf{E})~Tomographic fidelities of the five measurements employed for the quantum information tasks. In each case, the device is re-calibrated to optimize the implementation. Each value is obtained by averaging over more than 10 million photon counts, and the error bar represents the standard deviation calculated from repeated experiments. 
    PMs: projective measurements; USD: unambiguous state dicrimination; SE: state estimation; RNG: random number generation; MMI BS: multimode interference beam splitter.
    } 
\end{figure*} 

Compared to general quantum measurements, projective measurements are simpler to implement since they require no interaction between the system and an ancilla. Achieving such general system-ancilla control is essential for implementing the full range of quantum measurements, posing a significant challenge for developing universal photonic measurement devices. 
The challenge is important to overcome because it is well-known that reaching beyond projective measurements is crucial for quantum information applications. Prominent examples include  tomography \cite{Bent15Experimental, Nguyen22Optimizing}, state discrimination \cite{barnett09quantum,Bergou12Opt} and cryptography \cite{Acin16Optimal, farkas24maximal}.
Driven by the goal of unlocking the full potential of quantum measurements, photonic experiments have reported many realizations, primarily in bulk optics \cite{Bian15Realization,Tavakoli20, martinez23certification, hou2018deterministic, Wang23Generalized, Goel23Simultaneously} and recently also in integrated circuits \cite{Feng25Integrated}. However, these are mostly single- or few-purpose devices, tailored for realizing one (or a specific class of) quantum measurement. Furthermore, the fidelity and setup complexity of these measurements vary significantly across platforms.

Here, we report a high-performance fully programmable measurement device based on integrated photonic technology. 
{It transforms photonic quantum measurements, up to the circuit size, from being a diverse collection of specialized tools to being defined by the programming of a single universal quantum device}
, as illustrated in \fref{fig: setup}A.  
The fabricated device integrates more than 700 components, including 96 thermo-optic phase shifters, which provide programmability to implement arbitrary four-dimensional quantum measurements on path-encoded states.
Our system actively controls the ancillary quantum systems required to synthesize non-orthogonal measurement outcomes. 
We demonstrate the versatility of the device by tomographically reconstructing 100 random quantum measurements, achieving an average fidelity of 97.7\%. To benchmark its performance in concrete quantum information tasks, we deploy it for three well-established primitives in which projective measurements fall short of the full capability of quantum measurements: the unambiguous discrimination of quantum states, the estimation of an unknown state from multiple copies, and secure generation of random numbers. In all three tasks, our device outperforms what is theoretically possible with projective measurements, corresponding to 23 times lower error, 10.6\% higher estimation fidelity, and  37\% more randomness yield.
Our work demonstrates how general quantum measurements can be implemented with high quality on a single device based on scalable integrated photonic technology.

\bigskip
\noindent\textbf{Results}\\
\noindent\textbf{The measurement device} \\
\noindent When post-measurement quantum states are not a concern, a quantum measurement can be characterized by a set of operators $\{E_i\}$ that are positive semidefinite ($E_i\succeq 0$) and complete ($\sum_i E_i=\id$). It is well-known that any  quantum measurement acting on a $d$-dimensional system can be reduced to measurements that have at most $d^2$  possible outcomes \cite{Ariano2005}. This contrasts with  projective measurements, a sub-class of quantum measurements that support no more than $d$ possible outcomes.  When applied to a quantum state, $\rho$, the probability of obtaining the $i$-th outcome is given by the Born's rule: $p_i=\Tr\left(E_i\rho\right)$.

While many schemes exist for implementing quantum measurements beyond projective limitation, quantum walks provide a particularly successful approach \cite{Kurzy13Quantum}. This method has been used in bulk optics to realize various measurements \cite{Bian15Realization, hou2018deterministic, Wang23Generalized}. We adapt this approach to an integrated photonic platform, in which individual  control of the system-ancilla interaction enables a programmable measurement device capable of implementing arbitrary measurements. 

Our  circuit for measuring a system consisting of $d$ path modes, $\{\ket{k}\}_{k=0}^{d-1}$, relies on $d^2-1$ cascaded modules, as illustrated for $d=4$ in \fref{fig: setup}B. The $i$-th module is configured to realize the outcome associated with the theoretical measurement operator $E_i$. The interaction with the ancillary mode $\ket{d}$  is realized via $d$ reconfigurable two-dimensional unitary operators $C_i^j$ ($j=1,\dots,d$), each acting on two adjacent modes $\ket{j-1}$ and $\ket{j}$.
The algorithm for determining $C_i^j$ is detailed in Methods. The operators $C_i^j$ %for $j<d$ (resp.~$j=d$) 
are physically realized using reconfigurable two-mode Mach-Zehnder interferometers composed of two 50:50 beam splitters and two (resp.~one) thermal-optical phase shifters. To demonstrate the circuit scheme in \fref{fig: setup}B
we have fabricated a silicon integrated photonic device. The circuit for state preparation is also integrated on the chip.
A photograph and size comparison of the device appears in \fref{fig: setup}C. See the Supplemental Material for more details on the experimental setup.

\bigskip
\noindent\textbf{Tomography of a hundred measurements}\\
\noindent To demonstrate both the programmability and quality of the measurement device, we have benchmarked it by tomographically reconstructing 100 distinct four-dimensional quantum measurements.  The measurements were selected at random, divided into five sets of 20 measurements with 4, 7, 10, 13 and 16 outcomes, respectively. The randomization was made by selecting the appropriate number of rows from a randomly generated unitary matrix. For each of the 100 programmings of the measurement device, we probe it with a well-selected set of  tomographically complete quantum states. For convenience, we probe the measurement device with a coherent laser instead of single photons and record the output intensities \cite{barnett22single}. 
From the resulting outcome statistics, we can reconstruct the measurement by using standard maximum likelihood estimation \cite{Fiur01maximum}. 

From the measurement reconstruction, we evaluate its fidelity with the sampled theoretical description of the measurement (see Methods section).
The results are shown in \fref{fig: setup}D. Averaged over all 100 samples, the fidelity is $97.7\%$. Naturally, the fidelity decreases and its variance increases with the number of phase shifters and MZIs actively used in the device. 
However, even at the maximum of 16 outcomes, the  fidelity remains high, with an average of 97.0\%.   These results were obtained without using output feedback to optimize the control parameters for the specific measurement under consideration.

\begin{table}[t]
        \renewcommand{\arraystretch}{1.5}
	\centering
	\caption{\label{tab: comparison} Fidelities of symmetric informationally complete measurements in various experiments. QC: quantum computing platform.}
	\begin{tabular*}{1\linewidth}{@{\extracolsep{\fill}} c| c c c c}
		\hline  
         Work  & Dimension & Fidelity & Platform\\
		\hline
        
		\rcite{Stricker22Experimental} & 2 & N/A & Trapped-ion QC\\ 

		\rcite{Wang23Generalized} & 3 & 0.949 & Bulk optics \\
  
        \rcite{Feng25Integrated} & 3 & 0.965 & Integrated optics \\

        \rcite{Ivashkov24high} & 4 & 0.704 & Superconducting QC\\

        This work & 4 & 0.981 & Integrated optics \\
		\hline  
	\end{tabular*}
\end{table}

\begin{figure}[t]
	\centering	
	\includegraphics[width=\linewidth]{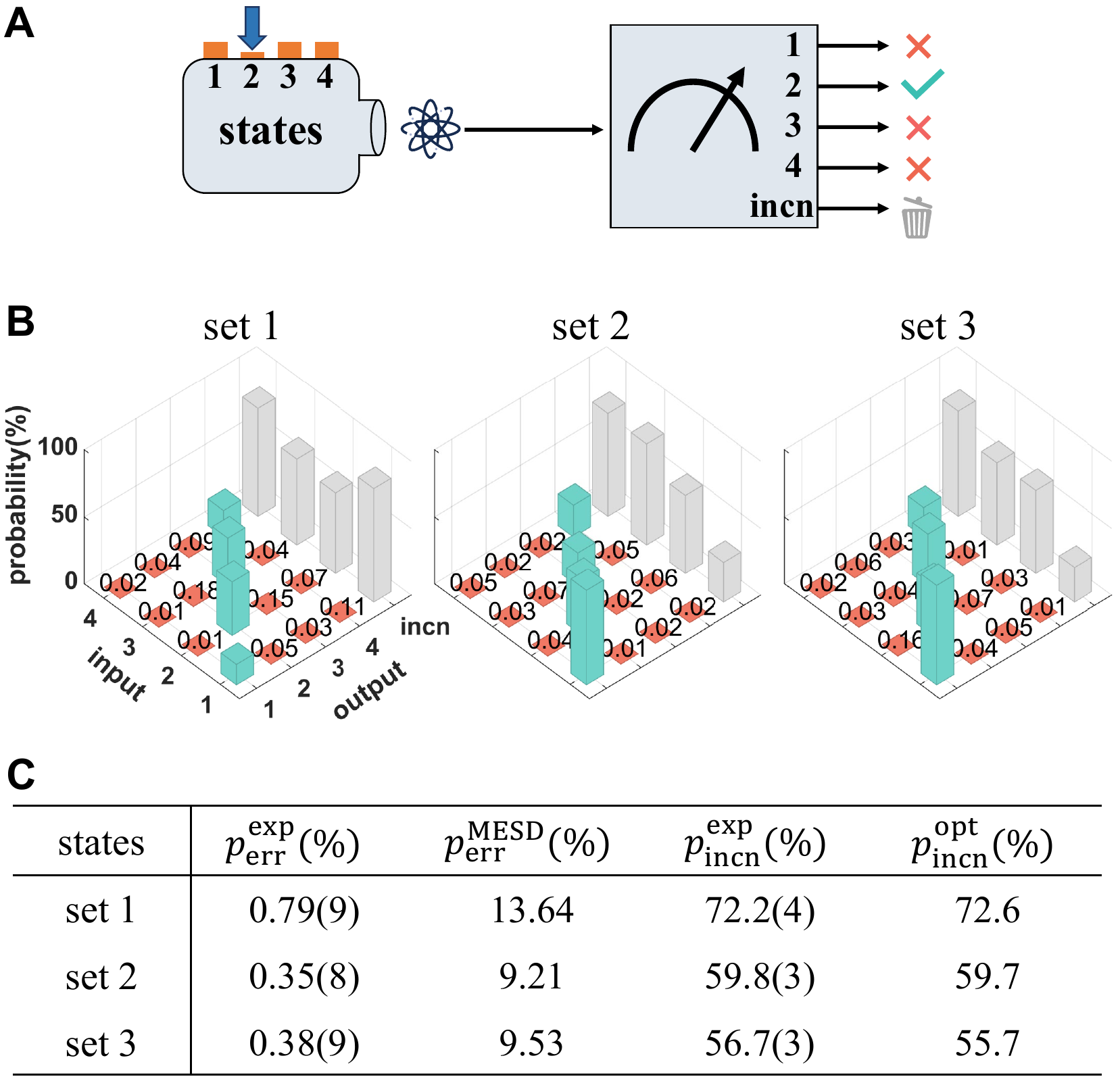}
	\caption{\label{fig: usd}  \textbf{Experimental  unambiguous state discrimination.} 
    (\textbf{A})~Schematic of unambiguously discriminating a set of four states  by allowing an inconclusive measurement outcome.
    (\textbf{B})~Experimental measurement probabilities for three different sets of states.
    The values for the incorrect outcomes are labeled on the corresponding bars. 
    (\textbf{C})~Experimental error rates compared with the theoretical limits of the minimum-error state discrimination protocol, and experimental probabilities of the inconclusive outcome compared with the theoretical optimal values. Each experimental value is the average over ten repeated experiments, with roughly $4\times4000$ photon counts collected in each, and the error bar indicates the standard deviation.} 
\end{figure}

\begin{figure*}[htpb]
	\centering	
	\includegraphics[width=\linewidth]{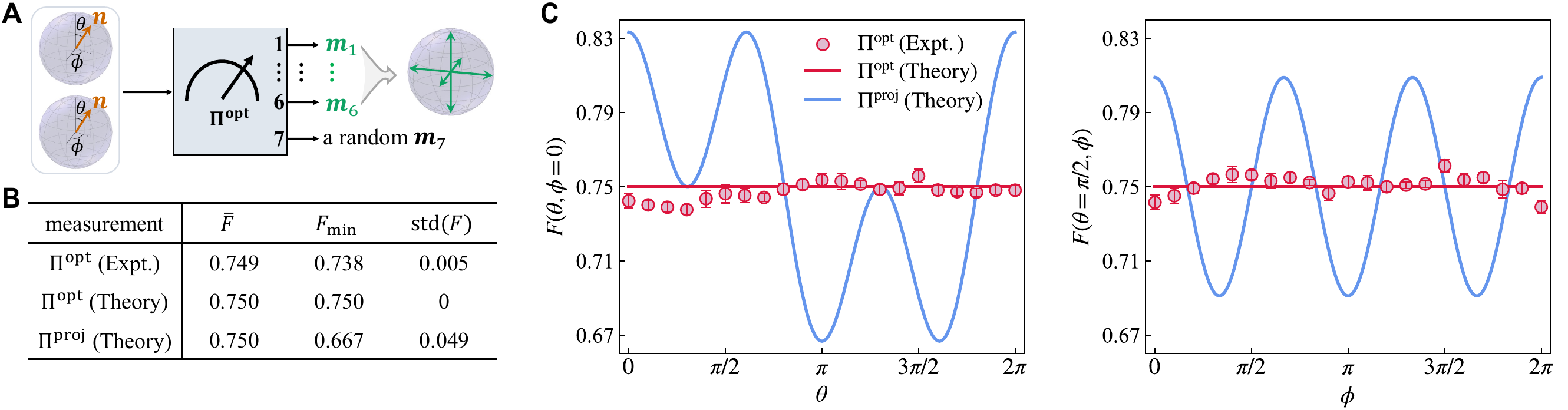}
	\caption{\label{fig: state estimation} \textbf{Experimental two-copy state estimation.} 
    (\textbf{A})~Schematic of the optimal estimation protocol. Two copies of an unknown pure qubit state are collectively measured and an estimate is constructed depending on the measurement outcome. 
    (\textbf{B})~Experimental estimation fidelities for the states with Bloch vectors on the $xz$ plane and $xy$ plane, respectively. Each data point is the average over ten repeated experiments, with roughly 4000 photon counts collected in each run, and the error bar indicates the standard deviation.
    (\textbf{C})~The average, minimum, and standard deviation of the experimental estimation fidelities across all input states studied, compared with the theoretical predictions and the corresponding results achieved by the projective measurement $\Pi^{\mathrm{proj}}$ proposed in \rcite{MassP95}.     } 
\end{figure*}

Then, we apply the measurement device to three distinct quantum information tasks: unambiguous state discrimination, multi-copy state estimation, and random number generation. In these tasks,  the best protocols rely on measurements beyond  projective limitations. Before performing each task, we optimize the fidelity of our measurement implementation by estimating and correcting the on-chip phase errors for each specific measurement (see the Supplemental Material).  With such a procedure, the fidelities can be increased as compared to those reported in \fref{fig: setup}D. 
This is showcased in \fref{fig: setup}E for the five measurements employed for the tasks. The results were obtained by performing measurement tomography with single-photon inputs. 
Furthermore, we note that these fidelities on many occasions exceed those reported from other platforms, even from experiments focusing on lower-dimensional measurements and using purpose-built measurement devices.
In Table~\ref{tab: comparison}  we provide a brief overview of the fidelities achieved for the seminal symmetric informationally-complete measurement \cite{renes2004symmetric,ZAUNER11}.

\bigskip
\noindent\textbf{Application I---unambiguous state discrimination}\\
\noindent Unambiguous state discrimination is a  well-known primitive in quantum information theory. The task is to make error-free identification of the classical label of a quantum state \cite{barnett09quantum}.
When the states are not orthogonal, this is made possible by introducing an additional outcome that is associated with an inconclusive result, meaning that the label could not be successfully identified. This is typically associated with implementing a measurement that is not projective \cite{ivanovic}. The  optimal measurement has been determined for arbitrary sets of linearly independent pure states \cite{Bergou12Opt}, but experiments venturing beyond qubit systems have so far been restricted to states with specific symmetries \cite{Agnew14Dis, Goel23Simultaneously}. Using our programmable device, we can realize the optimal protocol for a generic set of four-dimensional pure states. 

Consider four linearly independent four-dimensional states, $\{\ket{\Psi_k}\}_{k=1}^4$. Selecting one of them at random, we perform a measurement, $\{E_j\}_{j=1}^4\cup \{E_\text{incn}\}$, that aims to discriminate the state unambiguously, i.e.,~to output $j=k$. Our goal is for this discrimination to succeed with the highest possible average probability, while any failed discrimination is associated with the inconclusive outcome (instead of an incorrect result $j\neq k$). Under these conditions, we must therefore minimize the probability of the inconclusive outcome, $p_\mathrm{incn}=\frac{1}{4}\sum_k \bra{\Psi_k}E_\mathrm{incn}\ket{\Psi_k}$ (see the Supplemental Material for more details).
A schematic for the task is illustrated in \fref{fig: usd}A.

In the experiment, we made three separate  random selections for the set $\{\ket{\Psi_k}\}_{k=1}^4$ (the specific states are given in  Supplemental Material). 
\Fref{fig: usd}B displays the probabilities estimated from the experiment for each of the three sets of states. We  observe that the probability of incorrect identification ($j\neq k$), conditioned on a conclusive outcome, is small in all cases. In  \fref{fig: usd}C we compare the observed error rates with those achieved by projective measurements that only seek to minimize the error without an unambiguity constraint (so-called minimum-error state discrimination) \cite{Barnett_2009}. 
The error rate is reduced by a factor of 23 on average. Furthermore, the observed probability of the inconclusive outcome accurately matches the theoretical prediction; see \fref{fig: usd}C.

\bigskip
\noindent\textbf{Application II---multi-copy state estimation}\\
\noindent Another well-known task in quantum information processing is to estimate an initially unknown quantum state when multiple copies of it are available. It is known that collective measurements on two copies can provide a better estimation than individual measurements on each copy \cite{MassP95,Mass00collective,Tang2020Exp}. 
Here we consider the scenario in which we are given two copies of a random pure qubit state denoted by $\ket{\vec{n}}$, where $\vec{n}$ is its Bloch vector. We collectively measure both qubits and from the outcome $i$ we select an estimate,  $\vec{m}_i$, for the true Bloch vector $\vec{n}$. The estimation fidelity reads  
\begin{equation}\label{eq: fidelity0}
    F(\vec{n}) = \frac{1}{2}+\frac{1}{2}\sum_i p_i(\vec{n}) \ \vec{n}\cdot\vec{m}_i,
\end{equation}
where $p_i(\vec{n})$ is the probability of outcome $i$. Since $\vec{n}$ is unknown, one possible benchmark is to consider the average estimation fidelity over all possible $\vec{n}$, namely $\bar{F}=\int F(\vec{n}) \dd\vec{n}$. By virtue of a two-qubit projective measurement proposed in \rcite{MassP95}, denoted by $\Pi^{\proj}$ henceforth, 
we can construct an optimal protocol and achieve $\bar{F}=75\%$ \cite{MassP95}. When applying this protocol, however, some states are far worse estimated than the average. 

A more stringent benchmark is to consider the worst-case estimation fidelity,  $F_{\mathrm{min}}= \min _{\vec{n}} F(\vec{n})$. Interestingly, one can achieve $F=75\%$ for all $\vec{n}$ simultaneously, which means $F_{\mathrm{min}}=75\%$ (see the Supplemental Material for details). To this end, nevertheless, we need a seven-outcome measurement of the form $ \Pi_i^\opt=\frac{1}{2}\ketbra{\vec{m}_i}{\vec{m}_i}^{\otimes 2}$ for $ i=1,\dots,6$ and $\Pi_7^\opt=\id-\sum_{i=1}^{6}\Pi_i^{\mathrm{opt}}$, where $\{ \ket{\vec{m}_i}\}_{i=1}^6$ are the eigenstates of the three Pauli matrices. For the first six outcomes, we use $\vec{m}_i$ as our estimate, while for the final outcome we select an estimate at random. This protocol  is schematically illustrated in \fref{fig: state estimation}A.

In experiments, we implement the optimal measurement $\Pi_i^\opt$ and thereby  estimate states with Bloch vectors $\vec{n}=(\sin\theta \cos\phi, \sin\theta \sin\phi, \cos\theta)$ chosen first on the $xz$ plane ($\phi=0$) and then on the $xy$ plane ($\theta=\pi/2$). \Fref{fig: state estimation}B displays  the experimental estimation fidelities together with the theoretical predictions. For comparison, we also plot the theoretical results achieved by the projective measurement $\Pi^{\proj}$, which vary significantly across different input states. Compared with $\Pi^{\proj}$,  our experimental implementation of $\Pi^\opt$ can enhance the worst-case fidelity by 10.6\% and reduce the standard deviation across all input states by one order of magnitude, as shown in \fref{fig: state estimation}C.

\begin{figure}[t]
	\centering	
	\includegraphics[width=\linewidth]{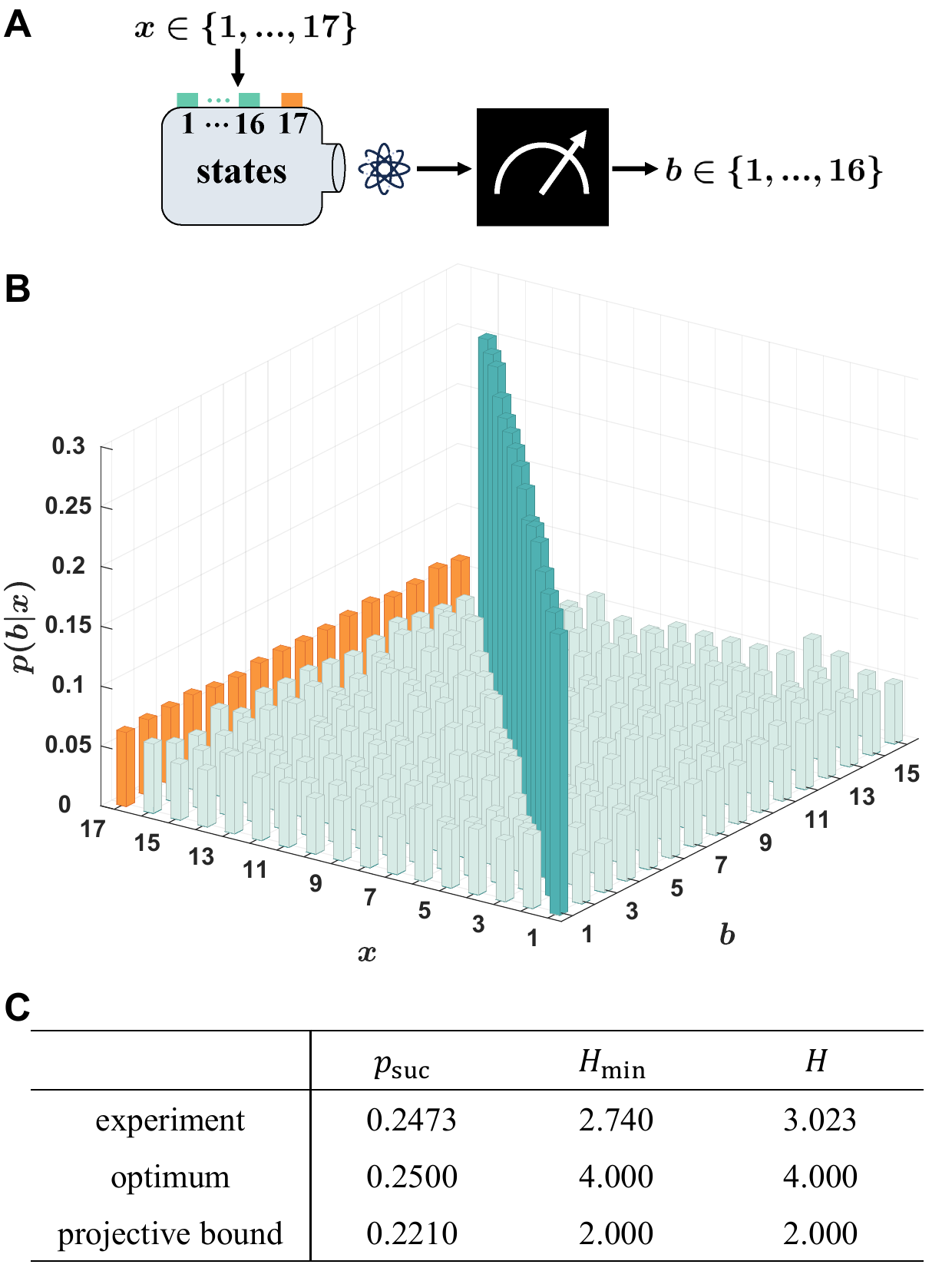}
	\caption{\label{fig: randomness}\textbf{Experimental  measurement-device-independent randomness certification.} 
    (\textbf{A})~Schematic of the protocol. 16 probe states are symmetric and informationally complete and one probe state is maximally mixed. The formers are used to ensure security and the latter is used to generate randomness from the output of the uncharacterised measurement.
    (\textbf{B})~Experimental measurement probabilities for the 17 input states.
    The results for inputs $x=1,\dots,16$ were obtained from roughly $16\times60000$ photon counts, while the results for the input $x=17$ were obtained from roughly 240000 photon counts.
    (\textbf{C})~Experimental success probability for the discrimination and lower bounds on min-entropy and Shannon entropy, compared with the optimal values and the corresponding bounds for projective measurements.}
\end{figure}

\bigskip
\noindent\textbf{Application III---random number generation}\\
\noindent Randomness is a resource for diverse applications, such as cryptography and simulation. Quantum mechanics predicts random numbers that  arise from the intrinsic randomness of measurement outcomes rather than from stochastic behaviors or ignorance. Their unpredictability can be certified from the point of view of an  eavesdropper with access to side-information \cite{mannalatha23}.
Here, we consider the measurement-device-independent approach to quantum random number generation \cite{bischof17measurement,supic17measurement}. This means the randomness is obtained by trusted states being used to probe a measurement device that is assumed to be under the eavesdropper's control. The latter is illustrated by a black-box in  the schematic of the scenario in  \fref{fig: randomness}A. 

If the measurement is four-dimensional and projective, then no more than  $\log_2(4)=2$ bits of randomness can be extracted per round since four is the largest number of possible outcomes.  By using a more general measurement, ideally with all 16 outcomes, this limit can be surpassed. To this end, we program our device to implement the seminal symmetric informationally complete measurement \cite{renes2004symmetric,ZAUNER11}. It formally corresponds to measurement operators  $E_i=\ketbra{\psi_i}{\psi_i}/d$ such that  $\left|\braket{\psi_j}{\psi_k} \right|^2=1/(d+1)$ for $j\neq k$. 
In our protocol, we probe the initially unknown measurement, $\{\Pi_i\}_{i=1}^{16}$, with the 16 states $\{\ket{\psi_i}\}_{i=1}^{16}$. By considering the discrimination probability $p_\text{suc}=\frac{1}{16}\sum_{i} \bra{\psi_i}\Pi_i\ket{\psi_i}$, we can certify how closely $\{\Pi_i\}_i$ approximates our target measurement. Then, we can use a 17'th state, selected as maximally mixed, to generate the randomness \cite{supic17measurement}. 

The probabilities estimated from the experiment are illustrated in \fref{fig: randomness}B. We see that the discrimination probability is high at all times: 0.2473 compared to the theoretical prediction of 0.25. This implies near-optimal security (see the Supplemental Material), which paves the way for us to use the statistics of the 17'th state for randomness generation. Moreover, the observed discrimination probability is on its own large enough to certify a genuine 14-outcome measurement, which is another signature of breaking projective limitations (see supplementary material). The observed probability distribution from the 17'th state is nearly uniform, which is desirable for high randomness generation. As detailed in the supplementary material, we obtain a lower bound on the certified randomness per round in the asymptotic limit, both in terms of the min-entropy and the Shannon entropy. These give  $H_\mathrm{min}=2.740$ and $H=3.023$,  respectively. We further perform a finite-size effects treatment and employ the entropy accumulation theorem \cite{metger2022}, which amounts to a rigorous lower-bound on the non-i.i.d.~certifiable randomness in the non-asymptotic limit. As a result, we finally obtain $2.979$ bits of randomness per-round. These results significantly surpass the corresponding bound achievable with projective measurements, as shown in \fref{fig: randomness}C.

\bigskip
\noindent\textbf{Discussion}\\
\noindent In summary, we have realized a high-quality programmable photonic circuit for universal quantum measurements. 
{On a single chip, we implemented 100 randomly selected four-dimensional measurements with a high average fidelity of 97.7\% and demonstrated three distinct quantum tasks with performance exceeding fundamental projective limits.} {Our work transitions quantum measurements from static or few-purpose devices to a dynamic, software-defined resource. Enabling such key functionality represents a significant step toward a full-stack integrated photonic quantum processor.}

Looking forward, several promising directions emerge from this work. First, our programmable measurement platform enables the rapid development of advanced measurement strategies—such as self-learning \cite{Rambach21Robust} and adaptive quantum measurement protocols \cite{MahlRDF13,tian2023minimumconsumption}. Second, while our current implementation operates on four-dimensional systems, the architecture is inherently scalable to higher dimensions.
Constructing a universal $d$-dimensional measurement device requires roughly $d^3$ two-dimensional unitary operators, offering lower circuit complexity than roughly $d^4/2$ two-dimensional unitary primitives required by the standard Naimark dilation method \cite{Reck94,Clements16}.
Third, this approach could be extended to multi-photon systems by combining with quantum joining \cite{vite13joining} that maps a multi-photon state into a high-dimensional single-photon system. Finally, the monolithic integration of our measurement processor with on-chip quantum light sources and state-manipulation circuits would realize a complete, self-contained photonic quantum processing chip.
Such an integrated platform would provide a universal testbed for implementing and verifying various complex quantum protocols, substantially advancing quantum information processing capabilities.

\bigskip

\clearpage

\noindent\textbf{MATERIALS AND METHODS}

\noindent\textbf{Algorithm for realizing a general quantum measurement}
	
\noindent Consider an arbitrary $d$-dimensional quantum measurement $\{E_i=a_i \ket{\psi_i}\bra{\psi_i}\}_{i=1}^n$ composed of $n$ rank-1 operators, where $\left|\braket{\psi_i}{\psi_i}\right|=1$ and $0< a_i \leq 1$. 
It can be realized by appropriately configuring a cascade of $n-1$ modules, each following the general structure shown (for $d=4$) in the inset of \fref{fig: setup}B.
The $i$-th module, which is configured to realize the measurement outcome associated with $E_i$, contains $d$ two-dimensional unitary operators $C_i^j$ ($j=1,\dots,d$), each acting on adjacent modes $\ket{j-1}$ and $\ket{j}$.
Here, the modes $\{\ket{k}\}_{k=0}^{d-1}$ constitute the $d$-dimensional system of interest, while the mode $\ket{d}$ acts as an ancilla. 
The completeness condition implies that the remaining ports of the ($n-1$)-th module yield the outcome for the last operator $E_n$.

Before presenting our algorithm for determining $C_i^j$, 
we define a set of evolution operators $\{K_i\}_i$ (acting solely on the $d$-dimensional system) such that for any initial state $\ket{\varphi}$, the (unnormalized) state input into the $i$-th module is $K_i\ket{\varphi}$.
The operators $K_i$ are also determined by the algorithm like $C_i^j$.
We denote by $l_i$ the effective dimension associated with $K_i$, which means that $K_i$ has nonzero entries only in its first $l_i$ rows.
Write $C_i^j$ in matrix form 
\begin{equation}
C_i^j=\begin{pmatrix}c_{00}^{i,j}&c_{01}^{i,j}\\c_{10}^{i,j}&c_{11}^{i,j}\end{pmatrix}.
\end{equation}
The four matrix elements can be further parameterized by the phase shifts of the corresponding MZI. 
For instance, an MZI with two phase shifters $\alpha^{i,j}$ and $\beta^{i,j}$ on the upper paths before and between two 50:50 beam splitters, respectively, yields the parameterization
\begin{equation}\label{eq: parameterization}
C_i^j=\rmi \rme^{\frac{\rmi\beta^{i,j}}{2}}\begin{pmatrix}\rme^{\rmi \alpha^{i,j}}\sin\frac{\beta^{i,j}}{2}&\cos\frac{\beta^{i,j}}{2}\\\rme^{\rmi \alpha^{i,j}}\cos\frac{\beta^{i,j}}{2}&-\sin\frac{\beta^{i,j}}{2}\end{pmatrix}.
\end{equation}

To realize the target measurement $\{E_i\}_{i=1}^n$, the parameters in $C_i^j$ ($i\leq n-1$) are set via the following algorithm, whose architecture is similar to the quantum-walk-based one proposed in \rcite{Bian15Realization}.\\[6pt]
\noindent Initialize the evolution operator $K_1=\id_d$ and the effective dimension $l_1=d$. For $i=1,\dots,n-1$, \\[4pt]

\begin{enumerate}
\item   Introduce a normalized ket
\begin{equation}
    \ket{\eta_i}={b^{-1}_i}(K_i^{+})^\dagger\ket{\psi_i},
\end{equation}
where $K_i^{+}$ is the Moore-Penrose generalized inverse of $K_i$ and $b_i:=\Vert (K_i^{+})^\dagger\ket{\psi_i}\Vert$. 

\item   Introduce the sub-stage evolution operators $\{R_i^j\}_{j=1}^{l_i}$ and initialize $R_i^1=\id_d$.

\item   If $l_i\geq 2$, 
introduce states $\{\ket{\xi_i^j}\}_{j=1}^{l_i}$ and initialize $\ket{\xi_i^1}=\ket{\eta_i}$.
For $j=1,\dots,l_i-1$, \\[4pt]
\indent (a) Choose $C_i^j$  such that
\begin{equation}\label{eq: condition1}		 
   \bra{j-1}U_i^j\ket{\xi_i^j}=0,
\end{equation}
 where 
   \begin{align}	
 \quad U_i^j=\ &c_{00}^{i,j}\ket{j-1}\bra{j-1}+c_{01}^{i,j}\ket{j-1}\bra{j}+c_{10}^{i,j}\ket{j}\bra{j-1}\nonumber\\
 &+c_{11}^{i,j}\ket{j}\bra{j} +\sum_{k\in\{0,1,\dots,d-1\} \setminus \{j-1,j\}}\ket{k}\bra{k}
   \end{align}	
   is the embedding of $C_i^j$ ($j \leq l_i-1$) into the $d$-dimensional system.  

\indent (b) Update $\ket{\xi_i^{j+1}}=U_i^j\ket{\xi_i^j}$ and $R_i^{j+1}=U_i^jR_i^j$.\\

\item   Choose $C_i^{l_i}$ such that
\begin{equation}\label{eq: condition2}
    \left|c_{10}^{i,l_i}\right|=b_i\sqrt{a_i}.
\end{equation} 

\item   If $l_i\leq d-1$, then, for $j=l_i+1,\dots,d$, choose $C_i^j$ with
\begin{equation}\label{eq: condition3}
    c_{00}^{i,j}=0.
\end{equation}

\item   Update the evolution operator as follows:
\begin{equation}
    K_{i+1}=\left(\sum_{k=0}^{l_i-2}\ket{k}\bra{k}+c_{00}^{i,l_i}\ket{l_i-1}\bra{l_i-1}\right)R_i^{l_i}K_{i},
\end{equation}

\item  If $c_{00}^{i,l_i}$=0, update $l_{i+1}=l_i-1$. Otherwise, update $l_{i+1}=l_i$. \\[6pt]

\end{enumerate}

\indent The algorithm works as follows: In the $i$-th module with effective dimension $l_i$, the sequence $\{C_i^j\}_{j=1}^{l_i-1}$ maps $\ket{\eta_i}$ into $\ket{l_i-1}$;
the subsequent $\{C_i^j\}_{j=l_i}^{d}$ diverts a fraction  of the amplitude $c_{10}^{i,l_i}$ from the mode $\ket{l_i-1}$ to the ancillary mode $\ket{d}$.
Therefore, the probability of an arbitrary probe state $\ket{\varphi}$ being detected at the $i$-th ($i\leq n-1$) output of the measurement device reads 
\begin{equation}
\left|c_{10}^{i,l_i}\bra{\eta_i} K_i\ket{\varphi}\right|^2=\Tr\left(\left|c_{10}^{i,l_i}\right|^2 K_i^\dagger\ketbra{\eta}{\eta}K_i\ketbra{\psi}{\psi}\right),
\end{equation} 
implying the realization of the measurement operator $\Pi_i=\left|c_{10}^{i,l_i}\right|^2 K_i^\dagger\ketbra{\eta}{\eta}K_i$. In the Supplemental Material, we prove that $\Pi_i=E_i$ for $i\leq n-1$ and that $c_{00}^{i,j}=0$ for $j > n-i$.

According to the specific parameterization in \eref{eq: parameterization}, \eref{eq: condition1}  determines both $\alpha^{i,j}$ and $\beta^{i,j}$ for $j=1,\dots, l_i-1$, whereas \eref{eq: condition2} or \eref{eq: condition3} determines only $\beta^{i,j}$ for $j= l_i,\dots, d$, in which case we can simply set $\alpha^{i,j}=0$. 
Any $d$-dimensional quantum measurement can be decomposed into extremal measurements that have at most $d^2$ rank-1 operators \cite{Ariano2005}. Hence, a cascade of $d^2-1$ modules suffices to construct a universal measurement device, as shown in \fref{fig: setup}B for $d=4$. 
In the last $d-1$ modules, the operators $C_i^j$ for $j>d^2-i$ are configured with $c_{00}^{i,j}=0$ regardless of the target measurement. Therefore, the detector for $E_i$ can be relocated to mode $\ket{d^2-i}$ directly after $C_i^{d^2-i}$.

\bigskip

\noindent\textbf{Definition of the measurement fidelity}

\noindent We evaluate the fidelity between the experimentally reconstructed measurement $\{E_i^\ex\}_{i=1}^n$ and the target measurement $\{E_i^\ideal\}_{i=1}^n$ using the same method as in \rcite{hou2018deterministic}. Specifically, we first map the two measurements to quantum states defined as $\sigma=\sum_{i=1}^n E_i^\ex \otimes\ketbra{i}{i}/d$ and $\sigma'=\sum_{i=1}^n E_i^\ideal \otimes\ketbra{i}{i}/d$, where $\{\ket{i}\}_{i=1}^n$ is an orthonormal basis and $d$ is the system dimension. The fidelity between the two measurements is then defined as the fidelity between their associated states $\sigma$ and $\sigma'$:
\begin{equation}
    F(\sigma, \sigma') := \left( \Tr \sqrt{\sqrt{\sigma} \sigma' \sqrt{\sigma}} \right)^2 = \left( \sum_{i=1}^n w_i \sqrt{F_i} \right)^2,
\end{equation}
where  $F_i=F\left(E_i^\ex/\Tr\left(E_i^\ex\right),  E_i^\ideal/\Tr\left(E_i^\ideal\right)\right)$ and $w_i=\sqrt{\vphantom{\frac{1}{1}} \Tr\left(E_i^{\ex}\right)\Tr\left(E_i^\ideal\right)}/d$.

\bibliography{all_references}

\bigskip
\noindent \textbf{Acknowledgments}\\
\noindent We thank Hefei Guizhen Chip Technologies Co., Ltd. for collaborating to develop the multichannel current-voltage source.

\noindent \textbf{Funding:} The work at the University of Science and Technology of China is supported by the Quantum Science and Technology-National Science and Technology Major Project (Grant Nos. 2024ZD0300900, 2023ZD0301400, 2021ZD0303200, and 2021ZD0301500), the National Natural Science Foundation of China (Grants Nos. 92576107, 62222512, 12104439, 12134014, T2325022, U23A2074, 6227524, and 62435009), CAS Project for Young Scientists in Basic Research (YSBR-049), the Anhui Provincial Natural Science Foundation (Grant No.2208085J03), and the Fundamental Research Funds for the Central Universities. 
The work at Quantum Science Center of Guangdong-Hong Kong-Macao Greater Bay Area is supported by the National Natural Science Foundation of China (Grant No. 12574401), the Guangdong Project (Grant No. 2024TQ08A680), the Guangdong Provincial Quantum Science Strategic Initiative (GDZX2403005, GDZX2403001, and GDZX2403002).
The work at Fudan University is supported by Shanghai Science and Technology Innovation Action Plan (Grant No.~24LZ1400200), Innovation Program for Quantum Science and Technology (Grant No. 2024ZD0300101), Shanghai Municipal Science and Technology Major Project (Grant No.~2019SHZDZX01), and the National Key Research and Development Program
of China (Grant No.~2022YFA1404204). CRC and AT are financially supported by the Wenner-Gren Foundations, by the Knut and Alice Wallenberg Foundation through the Wallenberg Center for Quantum Technology (WACQT), by the Swedish Research Council under Contract No.~2023-03498 and by the Swedish Foundation for
Strategic Research. 

\noindent \textbf{Author contributions:} GYX and XFR supervised and managed the project. ZH, YYZ conceived the project. WZY, LTF, YYZ and ZH designed the device. WZY and LTF built the experimental setup, calibrated the system and carried out the experiments. HZ, AT and CRC developed the theoretical framework of the protocols. WZY, ZH and CRC analyzed the experimental data with assistance of HZ, AT, ZH, GYX and GCG. All authors contributed to the writing of the manuscript.

\noindent\textbf{Competing interests:} The authors declare no competing interests.

\noindent\textbf{Data and materials availability:} All data needed to evaluate the conclusions in the paper are present in the paper and/or the Supplementary Materials. Additional data related to this paper may be requested from the authors.

\end{document}

\title{Supplemental Material for \\ A single programmable photonic circuit for universal quantum measurements}
\maketitle

\tableofcontents

\section{Proof of universality of the algorithm for realizing a general measurement}

\begin{figure}[htpb]
	\centering	
	\includegraphics[width=0.8\linewidth]{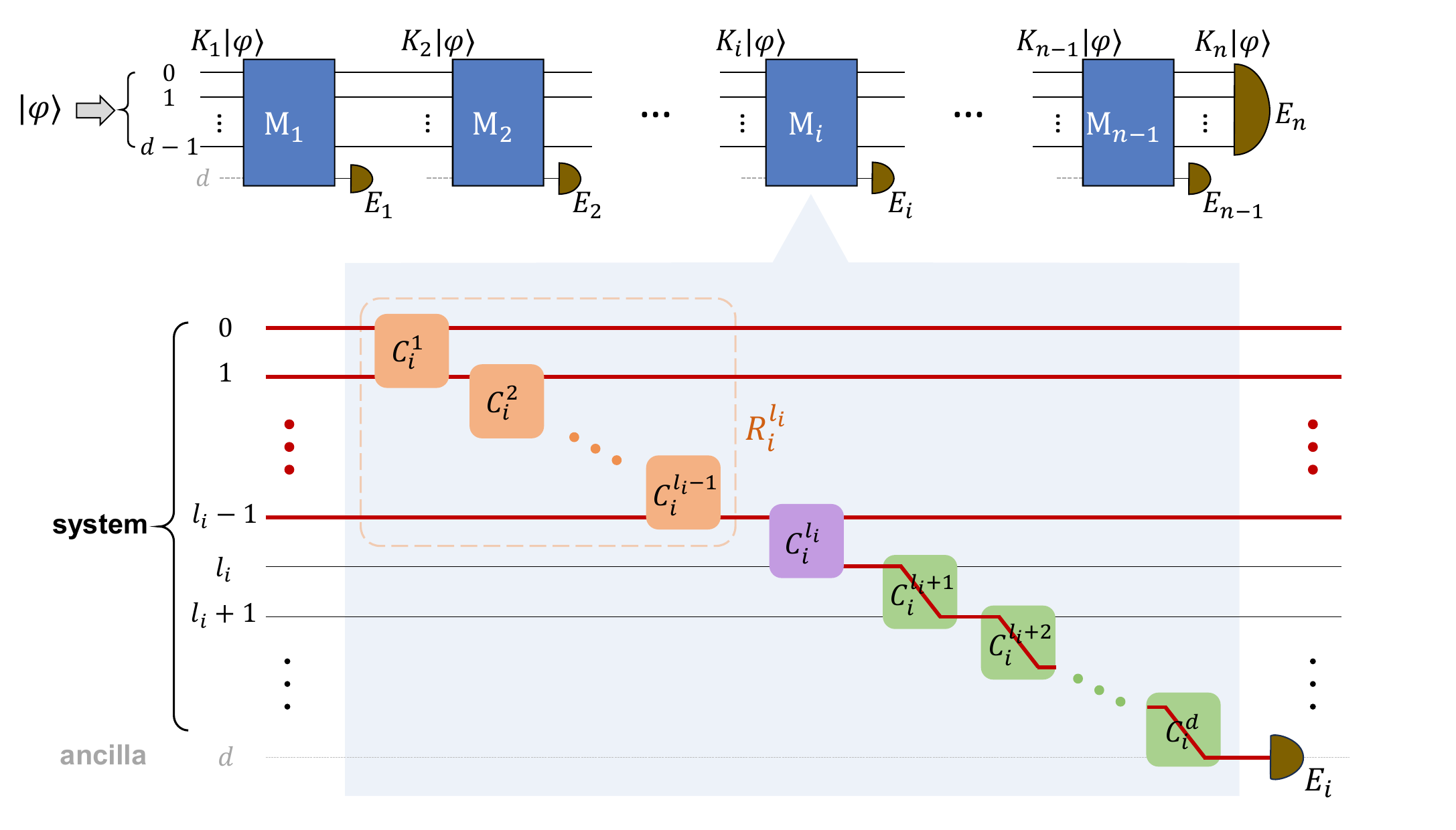}
	\caption{\label{fig: module} Schematic of the circuit for implementing an arbitrary $d$-dimensional measurement composed of $n$ rank-1 operators $\{E_i\}_{i=1}^n$. The modes $\{\ket{k}\}_{k=0}^{d-1}$ constitute the $d$-dimensional system of interest, and the mode $\ket{d}$ acts as an ancilla. The detector for $E_i (i<n)$ is placed at the ancillary mode $\ket{d}$ following the $i$-th module $\text{M}_i$. Here $\ket{\varphi}$ denotes an arbitrary initial state. The evolved state in the $i$-th module has nonzero components only in the modes highlighted with red lines.} 
\end{figure}

As described in the Methods section, an arbitrary $d$-dimensional measurement composed of $n$ rank-1 operators $\{E_i=a_i \ket{\psi_i}\bra{\psi_i}\}_{i=1}^n$, where $\ket{\psi_i}$ are normalized states and $0< a_i \leq 1$, can be realized by appropriately configuring a cascade of $n-1$ periodic modules following the circuit schematic in \Fref{fig: module}.
The $i$-th module $\mathrm{M}_i$ is configured to realize the measurement outcome associated with $E_i$. In the circuit, the two-dimensional unitary operators $C_i^j$, which have the form
\begin{equation}
C_i^j=\begin{pmatrix}c_{00}^{i,j}&c_{01}^{i,j}\\c_{10}^{i,j}&c_{11}^{i,j}\end{pmatrix},
\end{equation}
are determined by the algorithm in the Methods section, which we outline below:\\[12pt]
Introduce evolution operators $\{K_i\}_{i=1}^n$ and effective dimensions $\{l_i\}_{i=1}^n$. Initialize $K_1=\id_d$ and $l_1=d$. For $i=1,\dots,n-1$,\\[8pt]
1. Introduce a normalized ket $\ket{\eta_i}={b^{-1}_i}(K_i^{+})^\dagger\ket{\psi_i}$, where $b_i:=\Vert (K_i^{+})^\dagger\ket{\psi_i}\Vert$. \\[6pt]
2. Introduce sub-stage evolution operators $\{R_i^j\}_{j=1}^{l_i}$ and Initialize $R_i^1=\id_d$.\\[6pt]
3. If $l_i\geq 2$, 
introduce states $\{\ket{\xi_i^j}\}_{j=1}^{l_i}$ and initialize $\ket{\xi_i^1}=\ket{\eta_i}$.
For $j=1,\dots,l_i-1$, \\[4pt]
\indent (1) Choose $C_i^j$  such that $\bra{j-1}U_i^j\ket{\xi_i^j}=0$, where $U_i^j$ is the embedding of $C_i^j$ into the $d$-dimensional system.\\[4pt]
\indent (2) Update $\ket{\xi_i^{j+1}}=U_i^j\ket{\xi_i^j}$ and $R_i^{j+1}=U_i^jR_i^j$.\\[6pt]
4. Choose $C_i^{l_i}$ such that $\left|c_{10}^{i,l_i}\right|=b_i\sqrt{a_i}$.\\[6pt]
5. If $l_i\leq d-1$, then, for $j=l_i+1,\dots,d$, choose $C_i^j$ such that $c_{00}^{i,j}=0$.\\[6pt]
6. Update $K_{i+1}=\left(\sum_{k=0}^{l_i-2}\ket{k}\bra{k}+c_{00}^{i,l_i}\ket{l_i-1}\bra{l_i-1}\right)R_i^{l_i} K_{i}$.\\[6pt]
7. If $c_{00}^{i,l_i}$=0, update $l_{i+1}=l_i-1$. Otherwise, update $l_{i+1}=l_i$.\\[12pt]
\noindent The embedding of $C_i^j$ into the $d$-dimensional system is given by
\begin{equation}
   \begin{aligned}
   U_i^j&=c_{00}^{i,j}\ket{j-1}\bra{j-1}+c_{01}^{i,j}\ket{j-1}\bra{j}+c_{10}^{i,j}\ket{j}\bra{j-1}
   +c_{11}^{i,j}\ket{j}\bra{j}+\sum_{k\in\{0,1,\dots,d-1\} \setminus \{j-1,j\}}\ket{k}\bra{k} \quad \text{for} \quad j \leq d,\\
   U_i^d&=c_{00}^{i,d}\ket{d-1}\bra{d-1}+\sum_{k=0}^{d-2}\ket{k}\bra{k}.
   \end{aligned}
\end{equation}
$U_i^d$ has the above non-unitary form because $C_i^d$ couples the system to the previously unoccupied ancillary mode $\ket{d}$.
% The component in $\ket{d}$ is detected immediately after the application of  $C_i^d$.

The introduced operator $K_i$ is defined for describing the system evolution from the beginning to the input of the $i$-th module, i.e., for any initial state $\ket{\varphi}$, the (unnormalized) state input into the $i$-th module is given by $K_i\ket{\varphi}$. The introduced quantity $l_i$ is defined as the effective dimension associated with $K_i$, which means that $K_i$ has nonzero elements only in its first $l_i$ rows. It is easy to see that step 7 of the algorithm properly updates $l_{i+1}$. We now prove that step 6 properly updates $K_{i+1}$. 
According to step 2 and step 3(2),  $R_i^{l_i}$ (for $l_i\geq2$) represents the action of the sequentially applied $C_i^1,\dots,C_i^{l_i-1}$ on the 
$d$-dimensional system. 
According to step 4 and step 5, the action of sequentially applied $C_i^{l_i},\dots,C_i^{d}$ on the 
$d$-dimensional system reads
\begin{equation}
U_i^{l_i\rightarrow d}=c_{00}^{i,l_i}\ket{l_i-1}\bra{l_i-1}+c_{01}^{i,l_i}\ket{l_i-1}\bra{l_i}+\sum_{k=0}^{l_i-2}\ket{k}\bra{k}+\sum_{k=l_i+1}^{d-1}c_{01}^{i,k}\ketbra{k-1}{k},
\end{equation}
where $\left|c_{01}^{i,k}\right|=1$ for $l_i\leq k \leq d-1$.
% According to steps 2(5) and 2(6), $K_i$ has nonzero elements only in its first $l_i$ rows.
Since $K_i$ has nonzero elements only in its first $l_i$ rows and the nontrivial action of the unitary operator $\tilde{R_j}$ is confined to the subspace spanned by $\{\ket{k}\}_{k=0}^{l_i-1}$, we have 
\begin{equation} \label{eq: evolution}
\begin{aligned}
U_i^{l_i\rightarrow d}R_i^{l_i}K_i&=U_i^{l_i\rightarrow d}R_i^{l_i}\left(\sum_{k=0}^{l_i-1}\ketbra{k}{k}\right)K_i=U_i^{l_i\rightarrow d}\left(\sum_{k=0}^{l_i-1}\ketbra{k}{k}\right)R_i^{l_i}K_i\\
&=\left(\sum_{k=0}^{l_i-2}\ket{k}\bra{k}+c_{00}^{i,l_i}\ket{l_i-1}\bra{l_i-1}\right)R_i^{l_i}K_{i}.
\end{aligned}
\end{equation}
Because $U_i^{l_i\rightarrow d}\tilde{R_i}$ describes the system evolution within the $i$-th module,  step 6 properly updates $K_{i+1}$.
Note that the algorithm operates under the assumption $l_i \geq 1$. 
We will prove by induction that the first $j$ modules realize the first $j$ measurement operators for $j\leq n-1$, which guarantees (through the completeness condition) that the effective dimension $l_i$ cannot drop to zero before all measurement operators are realized.

We now prove the universality of the algorithm for realizing general measurements by analyzing the evolution of an arbitrary initial state $\ket{\varphi}$.
Since $K_i$ has nonzero elements only in its first $l_i$ rows, its Moore-Penrose generalized inverse $K_i^{+}$ has nonzero elements only in its first $l_i$ columns.
Consequently, the state $K_i\ket{\varphi}$ input into the $i$-th module and the normalized ket $\ket{\eta_i}$ defined in the algorithm both have nonzero components only in the modes $\{\ket{k}\}_{k=0}^{l_i-1}$.
If $l_i\geq2$, then according to step 3 of the algorithm, we have 
\begin{equation}
\bra{k}R_i^{l_i}\ket{\eta_i}=0 \quad \text{for} \quad k=0,\dots,l_i-2,
\end{equation}
which implies that the unitary operator $R_i^{l_i}$ transforms $\ket{\eta_i}$ to $\ket{l_i-1}$ (up to a global phase). 
If $l_i=1$, $\ket{\eta_i}$ is automatically identical to $\ket{l_i-1}$ (up to a global phase).
% It follows that the unitary operator $\tilde{R_i}$ transforms $\ket{\eta_i}$ to $\ket{l_i-1}$.
% Step 2.(2) of the algorithm implies that the unitary operator $\tilde{R_i}$ transforms $\ket{\eta_i}$ to $\ket{l_i-1}$ (up to a global phase).
Thus, immediately before the application of $C_i^{l_i-1}$, the amplitude of the evolved state in the mode $\ket{l_i-1}$ is given by
\begin{equation}
   \left|\bra{l_i-1} R_i^{l_i} K_i \ket{\varphi}\right|=\left| \bra{\eta_i} K_i\ket{\varphi}\right|,
\end{equation}
and the amplitude in modes $\{\ket{k}\}_{k=l_i}^{d}$ is 0.
Then, $C_i^{l_i}$ transforms $\ket{l_i-1}$ to $c_{00}^{i,l_i}\ket{l_i-1}+c_{10}^{i,l_i}\ket{l_i}$. If $l_i\leq d-1$, the combination of $C_i^{l_i+1},\dots,C_i^{d}$ transforms $\ket{l_i}$ to $\ket{d}$ (up to a global phase). 
Therefore, after applying $C_i^d$, the amplitude in the ancillary mode $\ket{d}$ reads $\left| c_{10}^{i,l_i}\bra{\eta_i} K_i\ket{\varphi}\right|$.
The probability of the detection event corresponding to the detector placed at mode $\ket{d}$ following the $i$-th module reads
\begin{equation}
    p_i=\left|c_{10}^{i,l_i}\bra{\eta_i} K_i\ket{\varphi}\right|^2=a_i\left|\bra{\psi_i} K_i^{+}K_i\ket{\varphi}\right|^2
    =a_i\left|\bra{\psi_i} P_{K_i} \ket{\varphi}\right|^2,
\end{equation}  
where $P_{K_i}=K_i^{+}K_i$ is the projector onto $\mathrm{supp}(K_i)$.
In the following, we prove by induction that $p_i=\Tr\left(E_i\ketbra{\varphi}{\varphi}\right)$ for $i=1,2,\dots, n-1$.

In the beginning, we have $P_{K_1}=K_1^{+}K_1=\id_d$, therefore,
\begin{equation}
    p_1=a_1\left|\bra{\psi_1} P_{K_1} \ket{\varphi}\right|^2=a_1\left|\braket{\psi_1}{\varphi}\right|^2=\Tr\left(E_1\ketbra{\varphi}{\varphi}\right),
\end{equation}  
implying that the first module implements $E_1$. Assuming that the first $j$  modules have implemented $E_1,\dots,E_j$, we have 
\begin{equation}\label{eq: K condition}
    K_{j+1}^{\dagger}K_{j+1}=\id_d-\sum_{i=1}^{j}E_i=\sum_{i=j+1}^{n}E_i\geq E_{j+1}=a_{j+1} \ket{\psi_{j+1}}\bra{\psi_{j+1}},
\end{equation} 
implying that $\ket{\psi_{j+1}}\in \mathrm{supp}(K_{j+1})$.
Therefore,
\begin{equation}
    p_{j+1}=a_{j+1}\left|\bra{\psi_{j+1}} P_{K_{j+1}} \ket{\varphi}\right|^2=a_{j+1}\left|\braket{\psi_{j+1}}{\varphi}\right|^2=\Tr\left(E_{j+1}\ketbra{\varphi}{\varphi}\right),
\end{equation}  
implying that the $(j+1)$-th module implements $E_{j+1}$. 
\Eref{eq: K condition} also implies that 
\begin{equation}
    b_{j+1}^2=\bra{\psi_{j+1}}K_{j+1}^{+}(K_{j+1}^{+})^{\dagger}\ket{\psi_{j+1}}\leq \frac{1}{a_{j+1}},
\end{equation} 
so $C_{j+1}^{l_i}$ is well defined with its parameter satisfying
\begin{equation}
\left|c_{10}^{j+1,l_i}\right|=b_{j+1}\sqrt{a_{j+1}} \leq 1.
\end{equation} 
By induction, $p_i=\Tr\left(E_i\ketbra{\varphi}{\varphi}\right)$ holds for $i=1,2,\dots, n-1$. The completeness condition implies that the remaining ports of the $(n-1)$-th module yield the outcome for $E_n$. In this way, the target measurement $\{E_i\}_{i=1}^{n}$ is indeed realized.

In the following, we prove that for any target measurement, we have $c_{00}^{i,j}=0$ whenever $j > n-i$.

First, we have
\begin{equation}
   \text{rank}(K_i)=l_i, \quad \text{for} \quad i=1,\dots,n,
\end{equation}
which can be proved by induction. In the beginning, we have $\text{rank}(K_1)=l_1=d$. Assume that $\text{rank}(K_j)=l_j$. The product $\tilde{R_j}K_j$ has nonzero elements only in its first $l_j$ rows and has rank $l_j$ because the nontrivial action of the unitary operator $\tilde{R_j}$ is confined to the subspace spanned by $\{\ket{k}\}_{k=0}^{l_i-1}$. Combining with step 6 and step 7 of the algorithm, we have $\text{rank}(K_{j+1})=l_{j+1}$.
Therefore, $\text{rank}(K_i)=l_i$ for $i=1,\dots, n$.
Then, we have 
\begin{equation}\label{eq: l_i}
l_i=\text{rank}(K_i^\dagger K_i)=\text{rank}\left(\sum_{m=i}^{n}E_m\right)\leq n-i+1.
\end{equation}
According to step 5 and step 7 of the algorithm, we have $c_{00}^{i,j}=0$ for $j\geq l_{i+1}+1$. Combining with \eref{eq: l_i}, we have
$c_{00}^{i,j}=0$ for $j > n-i$.

% According to Eqs.(9-11) in the main text, 
% % and combining the fact that $R_i^{l_i}$ is a unitary operator which acts nontrivially only on the first $l_i$ rows of $K_i$, 
% we have
% \begin{equation}
%    \text{rank}(K_i)=l_i.
% \end{equation}
% representing that the submatrix formed by the first $l_i$ rows of $K_i$ has full row rank.
% which can be proved by induction. When $i=1$, we have $\text{rank}(K_1)=l_1=d$. Assume $\text{rank}(K_j)=l_j$. Because $\tilde{R_j}$ is a unitary operator acting nontrivially only on an $l_j$-dimensional subspace, the product $\tilde{R_j}K_j$ has nonzero elements only in its first $l_j$ rows and has rank $l_j$. According to Eq.(9) and Eq.(11) in the main text, we have $\text{rank}(K_{j+1})=l_{j+1}$.
% Therefore, $\text{rank}(K_i)=l_i$ holds.

% The last $d-1$ modules can be further simplified by removing these trivial operators.
% Thus, in the last $d-1$ modules, the detector for $E_i$ can be moved to mode $\ket{n-i}$ directly after $C_i^{n-i}$. And the detector for $E_n$ can be placed at mode $\ket{0}$ following the $(n-1)$-th submodule.

\section{Details about the device and experimental setup} 

Based on the algorithm above, we develop a programmable large-scale quantum photonic integrated circuit in silicon to implement arbitrary four-dimensional quantum measurements. The whole chip (including the state preparation and measurement circuits) integrates 109 active thermal-optical phase shifters and more than 600 passive components, specifically including 308 thermal insulation slots, 128 multimode interference beam splitters (MMI BSs), 140 bonding pads, 15 directional couplers, and 55 grating couplers.

\begin{figure}[htpb]
	\centering	
	\includegraphics[width=\linewidth]{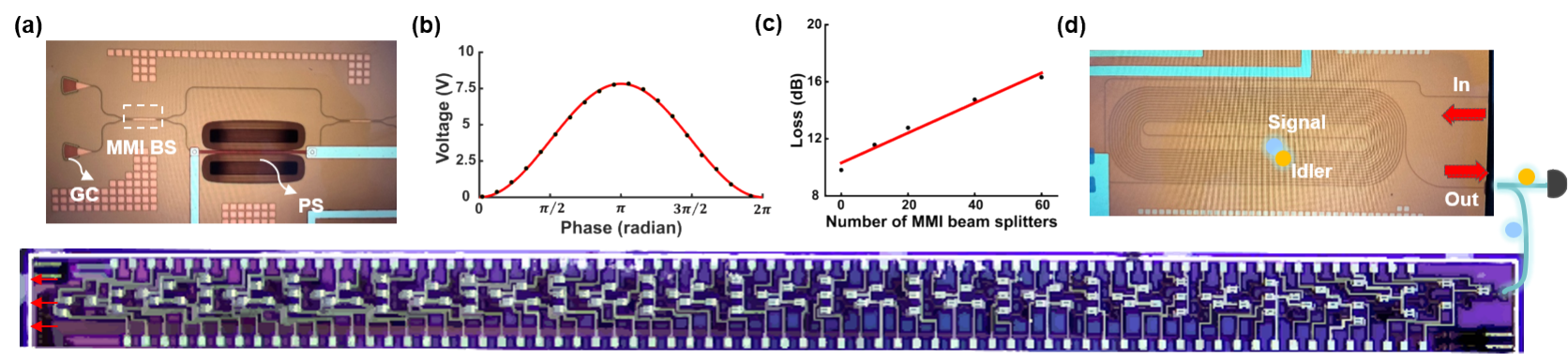}
	\caption{\label{fig: setupsm} (a) Micrograph of an on-chip Mach-Zenhder interferometer. GC: grating coupler; MMI BS: multimode interference beam splitter; PS: phase shifter. (b) One typical interference fringe of the phase shifter. The vertical axis represents the output voltage value of the photondetector, which is proportional to the optical power. The red fringe is obtained by fitting the data, and the visibility of the fringe is 99.9\%. (c) The loss results on cascaded MMI BSs with different quantities, and single MMI BS shows an insertion loss of about 0.1 dB. (d) Experimental setup for universal quantum measurements. The chip is driven by an integrated herald single-photon source. The chip size is 1.5$\times$15.6 mm$^2$ and the insertion loss is approximately 5.5 dB.} 
\end{figure}

The waveguide of the device is in a single mode and has a cross-section of 500$\times$220 nm$^2$. All phase shifters show resistance around 500 $\Omega$ and are controlled by the multichannel DC power supply, which delivers a stable electric current to each phase shifter individually. To avoid thermal crosstalk between phase shifters, two thermal insulation slots are placed on both sides of the phase shifter, which simultaneously reduces power consumption. The on-chip grating coupler is used for chip-fiber coupling with a coupling angle of 10$^\circ$, and the coupling loss is estimated to be approximately 4.5 dB. The MMI BS is used to divide the optical power equally into two parts. Two MMI BSs and one phase shifter constitute one Mach-Zenhder interferometer (MZI) for the programmable control of optical quantum states, as shown in \fref{fig: setupsm}(a). We calibrate each phase shifter with one continuous-wave laser at 1561.42 nm, and the average power consumption with 2$\pi$ phase modulation is less than 10 mW. The laser output is measured by one photodetector, and one example is given in \fref{fig: setupsm}(b). The data is fitted with the sinusoidal function with electrical power consumption. We compute the fringe visibility $V=(d_{\rm{max}}-d_{\rm{min}})⁄(d_{\rm{max}}-d_{\rm{min}})$ from the maximum $d_{\rm{max}}$ and minimum $d_{\rm{min}}$ of the fitted data. The average visibility of the fitted data is higher than 99\%. In the experiment, the chip is temperature controlled at 17.5 $^\circ$C with precision 0.001$^\circ$C, which is slightly lower than the ambient temperature for chip cooling. We measured the losses of cascaded MMI BSs with different quantities, as shown in \fref{fig: setupsm}(c). The loss of a single MMI BS is obtained by linear fitting.

The experimental setup is shown in \fref{fig: setupsm}(d). We employ one integrated herald single-photon source to drive the quantum measurement chip, and its core is a 1.2 cm long spiral silicon waveguide. The total loss of the packaged source from fiber input to fiber output is only 8 dB, including 3 dB transmission loss and 5 dB chip-fiber coupling input/output loss. A femtosecond erbium laser (T-light FC, pulse width 90 fs) with a repetition rate of 100 MHz is used as the pump source. After filtered by the 100 GHz bandwidth filter centering at 1550.12 nm, the left laser is amplified by an erbium-doped fiber amplifier and filtered again before being used to pump the waveguide. In the waveguide, the spontaneous four-wave mixing process occurs, and correlated photon pairs are generated. At the output end of the source, two cascaded off-chip post-filters (100 dB extinction ratio) are used to remove the pump photons first, then two wavelength division multiplexing filters with 200 GHz bandwidth are used to select signal and idler photons, which have respective central wavelengths of 1561.42 nm and 1538.98 nm. The photons are detected using superconducting nanowire single-photon detectors (SCONTEL, dark count rate 100 Hz, detector efficiency 85\% at C band). The electrical signals from the detectors were collected and analyzed through a time-correlated single-photon counting system, and the coincidence window was set at 1.28 ns. With only 100 $\mu$W pump power, the source shows a brightness of 100 kHz. 
In the benchmark of the programmable measurement chip, idler photons are directly input into the detector, whereas the heralded signal photons pass through the chip to undergo the state preparation and measurement process. 
% Further research could consider the integration of the photonic source, filters, and quantum measurement unit on a single chip to increase practicality.

\section{Quantum measurement tomography and chip fine-calibration}

\subsection{Quantum measurement tomography}
Quantum measurement tomography is a process to reconstruct an unknown quantum measurement from the measurement statistics obtained by probing it with a set of known quantum states.  We use this process to characterize the actually implemented measurement on the chip. 

We choose as the input states  20 tomographic-complete states $\{\ket{\psi_j^{\QMT}}\}_{j=1}^{20}$ corresponding to a complete set of mutually unbiased bases, given by the columns of the following five matrices \cite{BrieWB10}:
% \begin{equation}
% \begin{aligned}
% \id_4, \quad
% \frac{1}{2}\begin{pmatrix} -\rmi&\rmi&\rmi&-\rmi\\
%                 -\rmi&-\rmi&\rmi&\rmi\\
%                 1&-1&1&-1\\
%                 1&1&1&1\\\end{pmatrix}, \quad
% \frac{1}{2}\begin{pmatrix} -\rmi&\rmi&\rmi&-\rmi\\
%                 1&1&-1&-1\\
%                 -\rmi&\rmi&-\rmi&\rmi\\
%                 1&1&1&1\\\end{pmatrix}, \quad
% \frac{1}{2}\begin{pmatrix} 1&1&1&-1\\
%                 -1&1&1&1\\
%                 1&-1&1&1\\
%                 1&1&-1&1\\\end{pmatrix}, \quad
% \frac{1}{2}\begin{pmatrix} 1&-1&1&-1\\
%                 -\rmi&-\rmi&\rmi&\rmi\\
%                 -\rmi&\rmi&\rmi&-\rmi\\
%                 1&1&1&1\\\end{pmatrix}.
% \end{aligned}
% \end{equation}
\begin{equation}
\begin{aligned}
\id_4, \quad
\frac{1}{2}\begin{pmatrix} 1&1&1&1\\
                1&-1&1&-1\\
                i&\rmi&-\rmi&-\rmi\\
                i&-\rmi&-\rmi&\rmi\\\end{pmatrix}, \quad
\frac{1}{2}\begin{pmatrix} 1&1&1&1\\
                \rmi&-\rmi&\rmi&-\rmi\\
                1&1&-1&-1\\
                \rmi&-\rmi&-\rmi&\rmi\\\end{pmatrix}, \quad
\frac{1}{2}\begin{pmatrix} 1&1&1&1\\
                -1&1&1&-1\\
                1&-1&1&-1\\
                1&1&-1&-1\\\end{pmatrix}, \quad
\frac{1}{2}\begin{pmatrix} 1&1&1&1\\
                -\rmi&\rmi&\rmi&-\rmi\\
                -\rmi&-\rmi&\rmi&\rmi\\
                1&-1&1&-1\\\end{pmatrix}.
\end{aligned}
\end{equation}
We evaluate the reliability of the experimental state preparations by programming the first measurement module ($\mathrm{M_1}$ in \fref{fig: module}) to project each experimental preparation of $\ket{\psi_j^{\QMT}}$ onto a complete orthonormal basis that includes $\ket{\psi_j^{\QMT}}$ itself as one of the basis states. We obtain a high average preparation fidelity (defined as the experimental frequency of projecting the lab realization of $\ket{\psi_j^{\QMT}}$ onto the intended basis state $\ket{\psi_j^{\QMT}}$) of 99.5\% across the 20 input states.

% To ensure the reliability of the state preparations, we estimate the fidelity of each input state $\ket{\psi_j^{\QMT}}$.
% This is done by projecting the state onto an orthogonal basis comprising itself and its orthogonal complements, denoted by $\{ \ket{\psi_j^{\QMT}}, \ket{\psi_{j,\perp 1}^{\QMT}}, \ket{\psi_{j,\perp 2}^{\QMT}}, \ket{\psi_{j,\perp 3}^{\QMT}} \}$,
% Here we use only the first measurement module to perform each single basis-state projection sequentially.
% We inject a classical coherent laser into the chip and record the output intensities $\{I_0, I_1, I_2, I_3\}$ corresponding to the projections onto the respective basis states. The fidelity is then calculated as $I_0/(I_0+I_1+I_2+I_3)$. We experimentally obtained an average fidelity of 0.995 across the 20 input states, validating the reliability of the state preparation process.

Then we perform measurement tomography for 100 randomly sampled measurements. We program the measurement device to implement each target measurement $\{E_i^{\ideal}\}_i$ and probe it with the above 20 states $\{\ket{\psi_j^{\QMT}}\}_j$, obtaining experimental statistics $\{f_{ij}\}_{i,j}$. 
Here, $f_{ij}$ denotes the frequency of obtaining outcome $i$ when the input state is 
$\ket{\psi_j^{\mathrm{QMT}}}$. 
The theoretical prediction for $f_{ij}$ is given by $\Tr\left(E_i^{\ex}\ketbra{\psi_j^{\QMT}}{\psi_j^{\QMT}}\right)$, where $E_i^{\ex}$ is the experimental measurement operator.
We use the maximum likelihood method proposed in \rcite{Fiur01maximum} to reconstruct the set $\{E_i^{\ex}\}_i$. The reconstructed set is defined as the one that maximizes the likelihood function, i.e.,
\begin{equation}
\begin{aligned}
\{E_i^{\ex}\}_i:=&\ \text{arg } \underset{\{E_i\}_i}{\text{max }}\mathcal{L}, \quad \text{with }\mathcal{L}=\prod_i\prod_j\left[\Tr\left(E_i\ketbra{\psi_j^{\QMT}}{\psi_j^{\QMT}}\right)\right] ^{f_{ij}},\\
&\text{subject to} \quad E_i\succeq 0, \quad \sum_i E_i=\id.
\end{aligned}
\end{equation}
The fidelity between the reconstructed measurement and the target measurement is evaluated by 
\begin{equation}
F=\left(\sum_{i=1}^n w_i\sqrt{F_i}\right)^2,
\end{equation}
where 
\begin{equation}
F_i=\left (\Tr\left[\sqrt{\sqrt{\frac{E_i^{\ex}}{\Tr\left(E_i^{\ex}\right)}}\frac{E_i^\ideal}{\Tr\left(E_i^\ideal\right)}\sqrt{\frac{E_i^{\ex}}{\Tr\left(E_i^{\ex}\right)}}}\right]\right )^2
\end{equation} 
is the fidelity between the two normalized operators $E_i^{\ex}/\Tr\left(E_i^{\ex}\right)$ and $E_i^{\ideal}/\Tr\left(E_i^{\ideal}\right)$,
and
\begin{equation}
w_i=\frac{\sqrt{\Tr\left(E_i^{\ex}\right)\Tr\left(E_i^\ideal\right)}}{d}
\end{equation}
is a weight factor \cite{hou2018deterministic}.
Here we take dimension $d=4$. 

The estimated fidelities for the 100 randomly sampled measurements are shown in Fig.1d of the main text, with an average value of 97.7\%. The results are obtained using coherent-laser input instead of single-photon input.

\subsection{Chip fine-calibration for each specific measurement used in applications}

A primary factor limiting the measurement fidelity is the phase errors of the on-chip phase shifters, i.e., the actually implemented phases deviate from their ideal values. Through measurement tomography, we can not only characterize the actually implemented measurement but also estimate these phase errors. In the following, we present our method for estimating and correcting the phase errors for each specific measurement setting.

In the beginning, we set the on-chip phases as the ideal values $\{\phi_k^{\ideal}\}_k$ (determined by the algorithm given in the Methods section) to implement the target measurement $\{E_i^\ideal\}_i$.
Then we probe the measurement device with the 20 states $\{\ket{\psi_j^{\QMT}}\}_j$ and obtain a set of experimental frequencies $\{f_{ij}\}_{i,j}$.
The theoretical prediction for $f_{ij}$ can be derived in one of two distinct ways: \\
(1) From Born's rule: $\Tr\left(E_i^{\ex}\ketbra{\psi_j^{\QMT}}{\psi_j^{\QMT}}\right)$, which is a function of the experimental measurement operator $E_i^{\ex}$; \\
(2) From circuit evolution: 
$p_{ij}^\mathrm{evol}(\ket{\psi_j^{\QMT}}, \{\phi_k^{\ex}\}_k)$, defined as the probability of the input state $\ket{\psi_j^{\mathrm{QMT}}}$ evolving to the $i$-th output of the measurement circuit (Fig.1b in the main text).
This probability is a function of the experimental on-chip phases $\{\phi_k^{\ex}\}_k$.\\
Through the first derivation, we reconstruct $\{E_i^{\ex}\}_i$ using the standard maximum likelihood method and calculate the first-round measurement fidelity $F_1$. Through the second derivation, we estimate the phase errors by searching for small deviations $\{\dd\phi_k\}_k$ that make the  probabilities $\{p_{ij}^\mathrm{evol}(\ket{\psi_j^{\QMT}}, \{\phi_k^{\ideal}+\dd\phi_k\}_k)\}_{i,j}$ coincide with the experimental statistics $\{f_{ij}\}_{i,j}$ as well as possible. 
The resulting set $\{\dd\phi_k\}_k$ represents an estimate of the experimental phase errors.
Then we update the on-chip phase settings from $\{\phi_k^{\ideal}\}_k$ to $\{\phi_k^{\ideal}-\dd\phi_k\}_k$ to compensate for phase errors, and perform the second round of measurement tomography. The second-round fidelity $F_2$ is generally higher than the first-round fidelity $F_1$.

As an example to demonstrate the effectiveness of the above optimization process, we improve the fidelity of the first measurement used in the task of unambiguous state discrimination from 0.9863 to 0.9916. The error rate in the state discrimination experiment is also significantly reduced, from 1.72\% to 0.63\%. The experimental data during this calibration stage are obtained using coherent-laser input.

After the fine-calibration process, we perform a final round of measurement tomography using single-photon input. The estimated fidelities for the five measurements used in the  quantum information tasks are shown in Fig.1e of the main text, obtained from  1400028, 1485233, 1519548, 2134477, 3953463 accumulated photon counts, respectively.  Each fidelity value is the average over ten (for the first four measurements) or six (for the last measurement) repeated experiments, and the error bar represents the standard deviation.

\section{Unambiguous quantum state discrimination}

\subsection{Linear independence of the states} 

Unambiguous discrimination is possible if and only if the states are linearly independent~\cite{CHEFLES1998}. 
The linear independence of a set of pure states $\{\ket{\Psi_i}\}_i$ can be quantified by the determinant of the Gram matrix $G$, whose elements are given by $G_{i,j}=\braket{\Psi_i}{\Psi_j}$. Here we consider a set of four-dimensional states $\{\ket{\Psi_k}\}_{k=1}^{4}$. The Gram matrix is given by 
\begin{equation}
\begin{aligned}
G=\begin{pmatrix}1 & \braket{\Psi_1}{\Psi_2} & \braket{\Psi_1}{\Psi_3} & \braket{\Psi_1}{\Psi_4} \\ \braket{\Psi_2}{\Psi_1} & 1 & \braket{\Psi_2}{\Psi_3} & \braket{\Psi_2}{\Psi_4} \\\braket{\Psi_3}{\Psi_1} & \braket{\Psi_3}{\Psi_2} & 1 & \braket{\Psi_3}{\Psi_4} \\ \braket{\Psi_4}{\Psi_1} & \braket{\Psi_4}{\Psi_2} & \braket{\Psi_4}{\Psi_3} & 1 \end{pmatrix}.
\end{aligned}
\end{equation}

For any set of states, $\mathrm{det}(G)\geq0$, and the equality holds if and only if the states are linearly dependent~\cite{Greub1975}. Geometrically, $\sqrt{\mathrm{det}(G)}$ is the volume of the parallelepiped spanned by $\{\ket{\Psi_i}\}_i$.
Therefore, $\mathrm{det}(G)$ serves as a quantitative measure of the linear independence.

\subsection{Optimal measurements} 

Consider the case where the four linearly independent non-orthogonal four-dimensional states $\{\ket{\Psi_k}\}_{k=1}^{4}$ are generated with equal prior probabilities.  
We use a five-outcome non-projective measurement to unambiguously identify the four states. 
The measurement operators are given by
\begin{equation}\label{eq: povm_usd}
\begin{aligned}
    E_j=a_j\ketbra{\Phi_j}{\Phi_j}&, \quad 0\leq a_j\leq 1, \quad j=1,\dots,4, \\ 
    E_\incn&=\id-\sum_{j=1}^{4}E_j.
\end{aligned}
\end{equation}
Each conclusive outcome $j$ indicates that the input state is $\ket{\Psi_j}$, while the inconclusive outcome "incn" yields no identification. To ensure no errors, each $\ket{\Phi_j}$ is uniquely determined (up to a global phase) as a normalized state orthogonal to all $\ket{\Psi_k}$ for $k \neq j$. 
The non-orthogonality of $\{\ket{\Psi_j}\}_{j=1}^{4}$ implies non-orthogonality of $\{\ket{\Phi_j}\}_{j=1}^{4}$, necessitating the inconclusive operator $E_\incn$ for the completeness of the measurement.
We optimize the coefficients $a_j$ in Eq.\eqref{eq: povm_usd} to minimize the  probability of the inconclusive outcome, 
\begin{equation}\label{eq: failure probability}
    p_\incn=\frac{1}{4}\sum_j \bra{\Psi_j}E_\incn\ket{\Psi_j}=1-\frac{1}{4}\sum_j a_j\left|\braket{\Psi_j}{\Phi_j}\right|^2.
\end{equation}
This optimization can be formulated as the following semidefinite program:
\begin{equation}
\begin{aligned}
    \underset{\{a_j\}_j}{\text{maximize}} & \quad \frac{1}{4}\sum_j a_j\left|\braket{\Psi_j}{\Phi_j}\right|^2\\
    \text{subject to} & \quad \id-\sum_j a_j\ketbra{\Phi_j}{\Phi_j}\succeq0, \quad 0\leq a_j\leq 1.
\end{aligned}
\end{equation}

\subsection{Experimental settings} 

The three sets of input states we choose are given by the following matrices, with columns corresponding to $\ket{\Psi_1}$, $\ket{\Psi_2}$, $\ket{\Psi_3}$, $\ket{\Psi_4}$.  

\begin{equation}\label{eq: usd states}
\begin{aligned}
&\mathrm{set1}=\begin{pmatrix}-0.3717-0.3117\rmi&0.4394-0.2619\rmi&-0.1983-0.3443\rmi&-0.0325-0.1753\rmi\\
0.1096+0.5635\rmi&0.2859+0.5227\rmi&-0.0971-0.2818\rmi&-0.0403-0.4469\rmi\\
-0.2687-0.2008\rmi&0.4953-0.3402\rmi&0.2266+0.5127\rmi&-0.0579-0.3399\rmi\\
0.4649-0.3263\rmi&0.0964+0.1142\rmi&-0.4841+0.4525\rmi&-0.5207-0.6139\rmi\end{pmatrix},\\
&\mathrm{set2}=\begin{pmatrix}0.3963+0.4143\rmi&0.2242+0.1814\rmi&0.1670-0.5253\rmi&-0.2909-0.0835\rmi\\
0.1252-0.4273\rmi&-0.6517+0.1848\rmi&0.0179-0.5335\rmi&-0.3509-0.2353\rmi\\
-0.1291+0.4315\rmi&0.2118-0.4864\rmi&0.3334-0.0495\rmi&-0.4095+0.5014\rmi\\
0.3979-0.3345\rmi&0.4051-0.1118\rmi&0.4434+0.3177\rmi&0.2368+0.5048\rmi\end{pmatrix},\\
&\mathrm{set3}=\begin{pmatrix}-0.4383-0.5134\rmi&-0.3202+0.0620\rmi&0.4884-0.0513\rmi&0.4286+0.3428\rmi\\
-0.5238-0.0632\rmi&0.1704-0.8889\rmi&-0.2911+0.1488\rmi&-0.0686+0.4814\rmi\\
-0.0482+0.3753\rmi&0.2068-0.1038\rmi&0.5337+0.3368\rmi&-0.1992+0.3649\rmi\\
-0.2591+0.2359\rmi&-0.0940+0.1102\rmi&-0.4580-0.2096\rmi&-0.5134+0.1611\rmi\end{pmatrix}.
\end{aligned}
\end{equation}
The linear independence for each set of states is quantified by the determinant of the Gram matrix:
\begin{equation}
    \mathrm{det}(G_1)=0.3011, \quad \mathrm{det}(G_2)=0.4446, \quad \mathrm{det}(G_3)=0.4275.
\end{equation}
The optimal value of the probability $p_\incn$ for each set of states is given by 
\begin{equation}
    p_\incn^\opt(\mathrm{set1})=72.59\%, \quad p_\incn^\opt(\mathrm{set2})=59.74\%, \quad p_\incn^\opt(\mathrm{set3})=55.75\%.
\end{equation}

% The optimal solution $q^{\opt}$ for each set of states is
% \begin{equation}\label{eq: optimal q}
% \begin{aligned}
% &q^{\opt}(\mathrm{set1})=\begin{pmatrix}0.8444,& 0.6070,& 0.6461,& 0.8062\end{pmatrix},\\
% &q^{\opt}(\mathrm{set2})=\begin{pmatrix}0.2930,& 0.5759,& 0.7560,& 0.7646\end{pmatrix},\\
% &q^{\opt}(\mathrm{set3})=\begin{pmatrix}0.2355,& 0.6109,& 0.6158,& 0.7677\end{pmatrix}.
% \end{aligned}
% \end{equation}
% They are the numerical results of \eref{eq: internal}. To verify their optimality, we randomly sampled $10^6$ admissible points for $q=(q_1, q_2, q_3, q_4)$ which satisfy the constraint in \eref{eq: constrain}, and compare their corresponding $p_\incn$ values with that of $q^{\opt}$. The results are shown in \fref{fig: usd_random_samples}. 
% It can be seen that each $q^{\opt}$ provides a lower bound on $p_\incn$ among all sampled points, confirming its optimality.

% \begin{figure*}[htpb]
% 	\centering	
% 	\includegraphics[width=\linewidth]{supp_usd_random_samples.pdf}
% 	\caption{\label{fig: usd_random_samples} 
%     Values of the no-answer probability $p_\incn$ computed from $10^6$ random samples of admissible $q$, compared with the bound provided by the optimal solution $q^{\opt}$ in \eref{eq: optimal q}.}
% \end{figure*} 

\subsection{Comparison with the minimum-error quantum state discrimination}

Experimental imperfections result in nonzero error rates of 0.79\%, 0.35\%, and 0.38\% in the unambiguous discrimination of the three sets of states in \eref{eq: usd states}, respectively. To evaluate the performance of our measurements, we compare these experimental error rates with the theoretical limits of the minimum-error state discrimination (MESD) protocol.
In MESD, each measurement outcome identifies one of the possible states and the overall error probability is minimized. The minimum-error probability for discriminating among the four input states ${\ket{\Psi_j}}_{j=1}^{4}$ are given by
\begin{equation}
\begin{aligned}
    p_\mathrm{err}^{\MESD}= 1-\underset{\{E_j\}_j}{\text{maximize}} & \quad \frac{1}{4}\sum_j \Tr\left( E_j\ketbra{\Psi_j}{\Psi_j} \right)\\
    \text{subject to} & \quad \sum_j E_j =\id, \quad E_j \succeq0.
\end{aligned}
\end{equation}
By solving the semidefinite program for each set of states in \eref{eq: usd states}, we have
\begin{equation}
p_\mathrm{err}^{\MESD}(\mathrm{set1})= 13.64\%,\quad
p_\mathrm{err}^{\MESD}(\mathrm{set2})= 9.21\%,\quad
p_\mathrm{err}^{\MESD}(\mathrm{set3})= 9.53\%.
\end{equation}
These minimum-error probabilities are an order of magnitude higher than the observed error rates in our unambiguous state discrimination experiment.

\section{Two-copy state estimation}

\subsection{Estimation fidelity achieved by the non-projective measurement $\Pi^\opt$}

The seven-outcome non-projective measurement $\Pi^\opt$ is given by 
\begin{equation}
    \Pi_j^\opt=\frac{1}{2}\ketbra{\vec{m}_j}{\vec{m}_j}^{\otimes 2},\quad j=1,\dots,6,\quad \Pi_7^\opt=\id-\sum_{j=1}^{6}\Pi_j^\opt,
\end{equation}
where 
\begin{equation}
\begin{aligned}
    \ket{\vec{m}_1}=\ket{0},\quad &\ket{\vec{m}_2}=\ket{1},\\
    \ket{\vec{m}_3}=\frac{1}{\sqrt{2}}(\ket{0}+\ket{1}),\quad &\ket{\vec{m}_4}=\frac{1}{\sqrt{2}}(\ket{0}-\ket{1}),\\
    \ket{\vec{m}_5}=\frac{1}{\sqrt{2}}(\ket{0}+\rmi\ket{1}),\quad &\ket{\vec{m}_6}=\frac{1}{\sqrt{2}}(\ket{0}-\rmi\ket{1}),
\end{aligned}
\end{equation}
with Bloch vectors
\begin{equation}
\vec{m}_1=(0, 0, 1), \quad \vec{m}_2=(0, 0, -1), \quad \vec{m}_3=(1, 0, 0),  \quad \vec{m}_4=(-1, 0, 0), \quad \vec{m}_5=(0, 1, 0), \quad \vec{m}_6=(0, -1, 0).
\end{equation}
Since  $\Pi_7^\opt$ is the projection onto the antisymmetric subspace of the two-qubit system, we have $p_7(\vec{n})=0$ for any input two-copy state $\ket{\vec{n}}^{\otimes2}$. Therefore, the estimation fidelity $F(\vec{n})$ is given by
\begin{equation}
    \begin{aligned}
    F(\vec{n}) &= \sum_{j=1}^6 \frac{1 + \vec{n}\cdot\vec{m}_j}{2}p_j(\vec{n})=\sum_{j=1}^6 \frac{1 + \vec{n}\cdot\vec{m}_j}{2}\Tr\left[ \Pi_j^\opt\ket{\vec{n}}\bra{\vec{n}}^{\otimes 2}\right] \\
    &=\frac{1}{2} \sum_{j=1}^6 \frac{1 + \vec{n}\cdot\vec{m}_j}{2}\left|\braket{\vec{n}}{\vec{m}_j}\right|^4=\frac{1}{2} \sum_{j=1}^6 \left(\frac{1 + \vec{n}\cdot\vec{m}_j}{2}\right)^3.
    \end{aligned}
\end{equation}
Write $\vec{n}$ as $\vec{n}=(n_x, n_y, n_z)$ with $n_x^2+n_y^2+n_z^2=1$, we have 

\begin{equation}
    \begin{aligned}
    F(\vec{n}) &= \frac{1}{16}\left[(1+n_x)^3+(1-n_x)^3+(1+n_y)^3+(1-n_y)^3+(1+n_z)^3+(1-n_z)^3\right]\\
    &= \frac{1}{16}\left[6+6\left(n_x^2+n_y^2+n_z^2\right)\right]=\frac{3}{4}.
    \end{aligned}
\end{equation}

\subsection{Estimation fidelity achieved by the projective measurement $\Pi^\proj$}

The projective measurement $\Pi^\proj$ proposed by Massar and Popescu in \rcite{MassP95} is given by 
\begin{equation}
    \Pi^\proj_j=\ket{\Psi^\mathrm{MP}_j}\bra{\Psi^\mathrm{MP}_j}, \quad j=1,\dots,4,
\end{equation}
where
\begin{equation}
    \ket{\Psi^\mathrm{MP}_j}=\frac{1}{2}\ket{\Psi_-}+\frac{\sqrt{3}}{2}\ket{\vec{m'}_j}^{\otimes2}.
\end{equation}
Here, $\ket{\Psi_-}=(\ket{01}-\ket{10})/\sqrt{2}$ is the singlet state and the four directions $\{\vec{m'}_j\}_j$ form a regular tetrahedron on the Bloch sphere. We can choose
\begin{equation}
\begin{aligned}
\ket{\vec{m'}_1}=\ket{0}, \quad
\ket{\vec{m'}_2}=\frac{\rmi}{\sqrt{3}}\left(\ket{0}+\sqrt{2}\ket{1}\right), \quad \ket{\vec{m'}_3}=\frac{\rmi}{\sqrt{3}}\left(\ket{0}+\rme^{2\pi \rmi/3}\sqrt{2}\ket{1}\right), \quad
\ket{\vec{m'}_4}=\frac{\rmi}{\sqrt{3}}\left(-\ket{0}+\rme^{\pi \rmi/3}\sqrt{2}\ket{1}\right).
\end{aligned}
\end{equation}
For a given $\vec{n}=(n_x, n_y, n_z)$ with $n_x^2+n_y^2+n_z^2=1$, we have 
\begin{equation}
    \begin{aligned}
    F(\vec{n}) &= \sum_{j=1}^4 \frac{1 + \vec{n}\cdot\vec{m'}_j}{2}p_j(\vec{n}) =\sum_{j=1}^4 \frac{1 + \vec{n}\cdot\vec{m'}_j}{2}\Tr\left[\ketbra{\Psi^\mathrm{MP}_j}{\Psi^\mathrm{MP}_j}\left(\ket{\vec{n}}\bra{\vec{n}}^{\otimes 2}\right)\right]\\
    &=\frac{1}{24}\left(18 + \sqrt{2}n_x^3 - 3\sqrt{2}n_x n_y^2 - 3n_x ^2n_z - 3n_y^2 n_z + 2n_z^3\right).
    \end{aligned}
\end{equation}
For Haar-randomly distributed $\vec{n}$, the average fidelity and the worst-case fidelity are given by
\begin{equation}
    \bar{F}=\int F(\vec{n}) \dd\vec{n}=\frac{3}{4}, \quad F_{\mathrm{min}}= \min _{\vec{n}} F(\vec{n})=\frac{2}{3}.
\end{equation}
 
\section{Random number generation using the SIC-POVM}

\subsection{Construction of the SIC-POVM in dimension four}

A $d$-dimensional symmetric informationally complete positive operator valued measure (SIC-POVM) is composed of $d^2$ elements $E_i=\ketbra{\psi_i}{\psi_i}/d$ with equal pairwise overlaps $\left|\braket{\psi_j}{\psi_k} \right|^2=(\delta_{jk}+1)/(d+1)$. We use the method in Ref.~\cite{renes2004symmetric} to construct a SIC-POVM in dimension $d=4$. The normalized vectors $\{\ket{\psi_i}\}_{i=1}^{16}$ can be generated by
\begin{equation}
    \ket{\psi_i}=D_{jk}\ket{\psi}, \quad j,k=0,\dots,3,  \quad i=j\times4+k+1,
\end{equation}
where $\ket{\psi}$ is a fiducial state and $\{D_{jk}\}_{j,k=0}^{3}$ are displacement operators given by
\begin{equation}
   D_{jk}=\omega^{\frac{jk}{2}}\sum_{m=0}^{3}\omega^{jm}\ket{k\oplus m}\bra{m}, \quad \text{with } \omega=\rme^{\frac{\pi \rmi}{2}}.
\end{equation}
The symbol $\oplus$ denotes addition modulo 4.

We choose one of the fiducial states proposed in \rcite{renes2004symmetric}, given by

\begin{equation}
    \ket{\psi}=\begin{pmatrix}r_0\\r_+\rme^{\rmi\theta_+}\\r_1\rme^{\rmi\theta_1}\\r_-\rme^{\rmi\theta_-}\end{pmatrix},
\end{equation}
where
\begin{equation}
    r_0=\frac{\sqrt{1-\frac{1}{\sqrt{5}}}}{2\sqrt{2-\sqrt{2}}},\quad  r_1=\left(\sqrt{2}-1\right)r_0,\quad r_{\pm}=\frac{1}{2}\sqrt{1+\frac{1}{\sqrt{5}} \pm \sqrt{\frac{1}{5}+\frac{1}{\sqrt{5}}}},
\end{equation}
\begin{equation}
    \theta_+=\frac{a}{2}+\frac{b}{4}+\frac{\pi}{4},\quad \theta_1=\frac{\pi}{2},\quad  \theta_-=-\frac{a}{2}+\frac{b}{4}+\frac{\pi}{4},
\end{equation}
\begin{equation}
    a=\arccos\frac{2}{\sqrt{5+\sqrt{5}}},\quad b=\arcsin\frac{2}{\sqrt{5}}.
\end{equation}

% \begin{conjecture}
% For any dimension $d$, let $\{\ket{k}\}_{k=0}^{d-1}$ be an orthonormal basis, and define
% \begin{equation}
%     \omega=\rme^{\frac{2\pi \rmi}{d}}, \quad D_{jk}=\omega^{\frac{jk}{2}}\sum_{m=0}^{d-1}\omega^{jm}\ket{k\oplus m}\bra{m},
% \end{equation}
% where $\oplus$ denotes addition modulo d. Then there exists a normalized fiducial state $\ket{\phi}$ such that the set $\{D_{jk}\ket{\phi}\}_{j,k=1}^{d}$ constitute a SIC-POVM.
% \end{conjecture}
% In $d=4$, we choose one of the fiducial states proposed in Ref.~\cite{renes2004symmetric}, given by 

\subsection{Certification of the multi-outcome feature of the SIC-POVM}

A four-dimensional SIC-POVM has 16 outcomes, which is the largest possible for an extremal four-dimensional measurement~\cite{Ariano2005}. 
However, a lab realization of this measurement is subject to unavoidable imperfections. Here, we certify the lab measurement in terms of the smallest number of outcomes necessary to reconstruct it. In other words, we investigate whether it is possible to reproduce the statistics of the lab measurement, $\{E^{\ex}_x\}_{x=1}^{16}$, using only classical randomness and quantum measurements with at most $N$ outcomes for $N \in \{1,\dots,16\}$. 

Our certification is based on examining a state discrimination witness. The states to be discriminated are the 16 states associated with the individual elements in the SIC-POVM, $\{\ket{\psi_x}\}_{x=1}^{16}$. By performing a measurement, $\{E_x\}$, the success probability is given by
\begin{equation}
p_{\suc}=\frac{1}{16}\sum_{x=1}^{16}\sandwich{\psi_x}{E_x}{\psi_x}.
\end{equation}
Notice that the maximal value,  $p_{\mathrm{suc}}$ = 1/4, is achieved if and only if $\{E_x\}_x$ is the SIC-POVM, i.e.~$E_x=\frac{1}{4}\ketbra{\psi_x}{\psi_x}$.

Restricting $\{E_x\}_x$ to have at most $N$ outcomes, we determine the largest possible value of $p_\suc$  by computing a series of semidefinite programs \cite{Tavakoli24Semidefinite}.
% \textcolor{red}{cite}.
%https://journals.aps.org/rmp/abstract/10.1103/RevModPhys.96.045006
For a given $N$, there are $\binom{16}{N}$ different semidefinite programs, indexed by all the sets $T\subset\{1,\ldots,16\}$ such that $T$ has  cardinality $N$.
\begin{equation}
\begin{aligned}
		\underset{\{E_x\}_x}{\text{maximize}} & \quad \frac{1}{16}\sum_{x=1}^{16}\sandwich{\psi_x}{E_x}{\psi_x}  \\
		\text{subject to} & \quad \sum_x E_x=\id, \quad E_x\succeq 0, \quad E_x=0 \quad \text{if} \quad x \notin T.
\end{aligned}
\end{equation}
Once all these semidefinite programs have been evaluated, the largest value among them corresponds to the largest value of $p_\suc$ achievable with $N$-outcome measurements. For $N=1,\ldots,16$, the results are shown in \tref{tab: success probability}. Note that since $p_\suc$ is linear in $E_x$, classical randomness does not play a role in this analysis. In the experiment, we observed an average value of $p_{\suc}^{\ex}=0.24730$ which corresponds to a 14-outcome measurements, i.e., the implemented measurement cannot be simulated by stochastically combining quantum measurements with at most 13 outcomes. 

% \textcolor{red}{Armin: Before it said we show 13-outcome but from the table it looks like it is 14 outcome. If you agree, please correct this everywhere.} \wenzhe{Sorry, I got it wrong. It is indeed 14-outcome.}

\begin{table}[htpb]
    \renewcommand{\arraystretch}{1.3}
    \caption{ \label{tab: success probability} The maximum value of the success probability $p_\suc$ in state discrimination achievable by using quantum measurements with at most $N$ outcomes.}
    \centering
    \begin{tabular}{p{3.35cm}<{\centering} | p{1.65cm}<{\centering} p{1.65cm}<{\centering} p{1.65cm}<{\centering} p{1.65cm}<{\centering} p{1.65cm}<{\centering} p{1.65cm}<{\centering} p{1.65cm}<{\centering} p{1.65cm}<{\centering}}  
    \hline
		 $N$ & 1 & 2 & 3 & 4 & 5 & 6 & 7 & 8 \\ 
      maximum $p_\suc$ & 0.0625 & 0.1184 & 0.1708 & 0.2210 & 0.2263 & 0.2323 & 0.2367 & 0.2392\\ 
    \hline
      $N$ & 9 & 10 & 11 & 12 & 13 & 14 & 15 & 16 \\ 
      maximum $p_\suc$ & 0.2418 & 0.2431 & 0.2445 & 0.2458 & 0.2471 & 0.2481 & 0.2491 & 0.2500\\    
    \hline
    \end{tabular}
\end{table}
% \wenzhe{Please check the decimals in the table. In your previous notes the discrimination metric is defined by $S=\frac{1}{4}\sum_{x=1}^{16}\sandwich{\psi_x}{E_x}{\psi_x}$, so I divide the  values given there by 4. But I am not sure if there are some rounding errors.}

\subsection{Measurement-device-independent randomness certification}

Here we present the methods used to compute bounds on the Shannon and min-entropies used to certify the produced randomness in the measurement-device-independent framework. Namely, we consider prepare-and-measure scenarios with fully characterised $d$-dimensional state preparations $\rho_x$, labelled by the classical input $x$, and unknown measurements. We certify the randomness produced in the measurement outcome with respect to a malicious eavesdropper, namely Eve. Eve is allowed to pre-program the measurement device through classical correlations labelled by $\lambda$. That is, the measurement device will perform a measurement strategy $E_{b}^\lambda$ with probability $q(\lambda)$ and produce a measurement outcome $b$. The only physical assumptions taken throughout the rest of this section are that state preparations $\rho_x$ are  trusted. 

Concretely, we use the success probability in state discrimination as a witness that acts as a security parameter, namely 
\begin{align} \label{eq:sd_suc}
W = \frac{1}{16} \sum_{x=1}^{16} \sum_\lambda q(\lambda) \Tr\left[\rho_x E_{x}^{\lambda} \right] \ ,
\end{align}
Note that this uses 16 out of the 17 states appearing in the randomness generation protocol. The formers are the $16$ four-dimensional SIC states and the latter is the maximally mixed state $\rho_{x^\ast}=\frac{\id}{4}$.
This is because the final state is not used for checking security but only for generating the randomness.

We begin presenting a method to bound the single-round min-entropy. We then introduce the method used to compute a min-tradeoff function used to bound the certifiable randomness through the entropy accumulation theorem. \\

%The methods we introduce here consist on semidefinite programming numerical tools. These are convex optimisation problems restricted to the cone of positive-semidefinite matrices. 

\textbf{Min-entropy.} To compute bounds on the min-entropy, we use the Leftover-Hash lemma, which states that the min-entropy can be expressed as the logarithm of Eve's guessing probability, i.e.~$H_\text{min}=-\log_2 p_g$. The guessing probability for a particular setting $x^\ast$ is
\begin{align}
	p_g = \sum_\lambda q(\lambda) \ \underset{b}{\max}\left\{ \Tr\left[ \rho_{x^\ast} E_{b}^\lambda \right] \right\} \ .
\end{align}
To optimise $p_g$ over all possible strategies $\lambda$, distributions $q(\lambda)$ and measurements $E_{b}^\lambda$ we perform two simplifications. First, we absorb the distribution $q(\lambda)$ into the POVM elements $E_{b}^\lambda$ and define a new operator $M_{b}^\lambda := q(\lambda) E_{b}^\lambda$. Secondly, we reduce the number of strategies $\lambda$ to only the relevant ones that yield maximal correlations \cite{Bancal2014}. That is, we consider $\lambda=b$ to be the strategy that yields the maximal correlations for $b$. Overall, we can write the final optimisation through the following semidefinite program,
\begin{equation}\label{eq:pgsdp_primal}
\begin{aligned}
		\underset{M_{b}^{\lambda}}{\text{maximize}} & \quad \sum_{\lambda}\Tr\left[\rho_{x^\ast} M_{\lambda}^{\lambda}\right]  \\
		\text{subject to} & \quad M_{b}^\lambda \succeq 0,  \\
		& \quad \sum_b M_{b}^\lambda = \frac{1}{d}\Tr\left[\sum_b M_{b}^\lambda\right]\id,  \\
		& \quad \sum_{x} \frac{1}{16} \sum_\lambda \Tr\left[ \rho_x M_{x}^\lambda \right] \geq  W  ,
\end{aligned}
\end{equation}
where $W$ is the state discrimination success probability from ~\eref{eq:sd_suc} used as the observable linear witness in the experiment. Any set of $d\times d$ matrices $M_{b}^\lambda$ that satisfy the constraints above represents a valid bound on the guessing probability, and consequently to the min-entropy. \\

\textbf{Shannon entropy.} We continue presenting a method to compute a min-tradeoff function $f_\text{min}$ to be used to bound the certifiable randomness through the entropy accumulation theorem \cite{metger2022}. Namely, 
\begin{align}
\label{eq:GEAT}
    \frac{1}{N}H_{\text{min}}^{\varepsilon}(B^{N}|E^{N}) \geq \underset{c_{N}\in\Omega}{\min} f_\text{min}\left(\text{freq.}(c_{N})\right) - \mathcal{O}(1/\sqrt{N}) \ ,
\end{align}
where $H_{\text{min}}^{\varepsilon}(B^{N}|E^{N})$ is the conditional smooth min-entropy, and $N$ is the total number of prepare-and-measure experimental rounds. A min-tradeoff function is defined as an affine function on the observable frequencies freq.$(c_N)$ that is lower-bounded by the von-Neumann entropy. The latter, in our case, reduces to the Shannon entropy under the assumption of classical side information. In order to bound $H_{\text{min}}^{\varepsilon}(B^{N}|E^{N})$, we therefore need to find a suitable affine function on the observable frequencies that lower-bounds the Shannon entropy. We will do so, by formulating the optimisation as a semidefinite program and finding its dual formulation, such that the constrained correlations are moved to the objective function, i.e., the min-tradeoff function we seek.  

We begin introducing a semidefinite program to bound the Shannon entropy. The conditional Shannon entropy on the measurement outcome can be written as
\begin{align}
	H = -\sum_\lambda q(\lambda) \sum_b p_{\lambda}(b|x^\ast) \log_2 p_{\lambda}(b|x^\ast), 
\end{align}
for $p_{\lambda}(b|x) = \Tr\left[\rho_x E_{b}^\lambda\right]$. To find a lower-bound on the Shannon entropy, we first express the logarithm in an integral fractional expression, which we can then discretize employing the Gauss-Radau quadrature and linearise the problem through an SDP relaxation (see Refs.\cite{brown2024,Carceller25,carceller2024photon}). This results into 
\begin{align}
	H \geq c_m + \sum_{i=1}^{m-1}\sum_{b} \tau_i g_{i,b} \label{eq:shann_inf}
\end{align}
with
\begin{align}
	g_{i,b} := \underset{z_{i,b}}{\text{inf}} \left\{ p(b|x^\ast) \left[2z_{i,b} + (1-t_i) z_{i,b}^2 \right] + t_iz_{i,b}^2 \right\} \nonumber
\end{align}
for $\tau_i := \frac{\omega_i}{t_i \log 2}$ and $c_m:=\sum_{i=0}^{m-1}\tau_i$, where $\omega_i$ and $t_i$ are the weights and nodes of the Gauss-Radau quadrature, $z_{i,b}$ are arbitrary scalar variables. We now compute the lower bound on the right-hand-side of \eqref{eq:shann_inf} using a semidefinite program relaxation. In order to do so, we will sample a list of operators $O=\{E_{b}, z_{i,a} E_{b} , z_{i,a}^2 E_{b} \ldots \}$ up to a certain order on $z_{i,a}$. With these operators we build the following block-matrix
\begin{align}
	G^{ia}_{b} = \begin{pmatrix}
	E_{b} & z_{i,a} E_{b} & z_{i,a}^2 E_{b} & \cdots \\
	z_{i,a} E_{b} & z_{i,a}^2 E_{b} & z_{i,a}^3 E_{b} & \cdots \\
	\vdots & \vdots & \vdots & \ddots
	\end{pmatrix} \succeq 0 \ ,
\end{align}
with monomials up to order $k$ in $z_{i,a}$, which is positive-semidefinite by construction. Moreover, note that each block $\left(G^{ia}_{b}\right)_{u,v} = z_{i,a}^{u+v} E_{b}$ for $u+v>0$ satisfies
\begin{align}
	&\sum_b \left(G^{ia}_{b}\right)_{u,v} = \frac{1}{d}\Tr\left[ \sum_b \left(G^{ia}_{b}\right)_{u,v} \right] \id \ ,
\end{align}
We can now re-write the Shannon entropy bound in \eqref{eq:shann_inf} and minimise over the defined operators above, first defining
\begin{align}
	&f_{i,a} = \sum_b \sum_{u,v} c_{u,v}^{i}(a,b) \Tr\left[\rho_{x^{*}} \left(G^{ia}_{b}\right)_{u,v} \right]
\end{align}
for 
\begin{align}
	c_{u,v}^{i}(a,b) = 2\delta_{b,a}\delta_{u+v,1}+(1-t_i)\delta_{b,a}\delta_{u+v,2} + t_i \delta_{u+v,2} \ .
\end{align}
Now, note we can render the minimisation of $f_{i,a}$ independently for each $i$. Therefore, we can omit the indexing $i$ on each variable, as it will be reset after each iteration. This leads to the following semidefinite program,
\begin{equation}\label{eq:shannon_primal} 
\begin{aligned}
    H \geq c_m + \sum_{i=1}^{m-1} \ \underset{G^{a}_{b}}{\text{minimize}} & \quad \tau_i \sum_{a,b} \sum_{u,v} c_{u,v}^{i}(a,b) \Tr\left[\rho_{x^{*}} \left(G^{a}_{b}\right)_{u,v} \right], \\
    \text{subject to} & \quad G^{a}_{b} \succeq 0, \quad \left(G^{a}_{b}\right)_{0,0} = E_{b}, \\
    & \quad \sum_{x} \frac{1}{16}\Tr\left[ \rho_x E_{x} \right] \geq W,  \\
    & \quad \sum_{b}E_{b} = \id , \\
    & \quad \sum_b \left(G^{a}_{b}\right)_{u,v} = \frac{1}{d}\Tr\left[ \sum_b \left(G^{a}_{b}\right)_{u,v} \right] \id .
\end{aligned}
\end{equation}
Any set of $kd\times kd$ matrices $G^{a}_{b}$ satisfying the primal constraints above represents a valid lower bound on the Shannon entropy. Here $2k$ is the order of the semidefinite program relaxation as $2k$ is defined as the maximum power of the scalars $z_{a}$ in the blocks of $G^{a}_{b}$.

\begin{table}
\begin{tabular}{|c|c|c|c|} \hline
	Order $(2k)$ & $m$ & H (Primal) & H (Dual) \\ \hline
	2 & 8 & 2.951 & 2.951 \\
	4 & 8 & 2.951 & 2.951 \\
	6 & 8 & 2.961 & 2.960 \\
	8 & 8 & 2.986 & 2.984 \\ 
    10 & 18 & 3.024 & 3.022 \\ \hline
\end{tabular}
\caption{Certifiable Shannon entropy using the state discrimination witness with the frequencies obtained from the prepare-and-measure experiment. We show the values obtained with the semidefinite program in its primal (\eqref{eq:shannon_primal}) and dual (\eqref{eq:shannon_dual}) formulations, given the Gauss-Radau quadrature $m$ and the order of the relaxation level $2k$.}
\label{tab:shannon}
\end{table}

We now proceed to derivate the dual form of the semidefinite program. The Lagrangian corresponding to the complete optimisation reads $\mathcal{L} = c_{m} + \sum_i \mathcal{L}_i$, for
\begin{equation}
\begin{aligned}
\mathcal{L}_i = & \sum_{a,b} \tau_i \left( 2\delta_{b,a}\delta_{u+v,1}+(1-t_i)\delta_{b,a}\delta_{u+v,2} + t_i \delta_{u+v,2} \right) \Tr\left[\rho_{x^{*}} \left(G^{a}_{b}\right)_{u,v} \right] - \sum_{a,b}\Tr\left[G^{a}_{b} \Gamma^{a}_{b}\right] \\
+ & s\left(\sum_{x} \frac{1}{16}\Tr\left[ \rho_x E_{x} \right] - W\right) + \Tr\left[R\left(\sum_{b}E_{b} - \id\right)\right] \\
+& \sum_{a,b}\sum_{u,v}\sum_{\omega=u+v}\Tr\left[Q_{\omega}^{a}\left(\left(G^{a}_{b}\right)_{u,v} - \frac{1}{d}\Tr\left[ \left(G^{a}_{b}\right)_{u,v} \right] \id\right)\right] ,
\end{aligned}
\end{equation}
where we introduced the dual variables $\Gamma^{a}_{b}$, $s$, $R$ and $Q_{\omega}^{a}$ as Lagrangian multipliers. In the last term we also took into account that the last constraint from the primal holds for all elements $\left(G^{a}_{b}\right)_{u,v}$ that satisfy $u+v=\omega$, $\forall \omega$ (i.e., all blocks with the same sum of indices $u$ and $v$ are equivalent, for $\omega$ denotes the power of the scalar $z_{a}$).

The solution to the original semidefinite program is always a saddle point of the Lagrangian $\mathcal{L}$, identified among the stationary points. The infimum of the Lagrangian over the variables of the primal semidefinite program reads
\begin{align}
\mathcal{I} = \underset{G^{a}_b}{\inf} \ \mathcal{L}  = c_m + \sum_{i=1}^{m-1} \underset{G^{a}_b}{\inf} \ \mathcal{L}_i  = c_m + \sum_{i=1}^{m-1} \underset{G^{a}_b}{\inf} \ \left\{\left(-sW - \Tr\left[R\right]\right) + \sum_{a,b} \Tr\left[G_b^{a}K_b^{a}\right]\right\} \ ,
\end{align}
for
\begin{equation}
\begin{aligned}
	\left(K_b^{a}\right)_{0,0} &= s\sum_{x}\frac{1}{16}\delta_{b,x}\rho_x + R - \sum_{a} \left(\Gamma_{b}^{a}\right)_{0,0} \\
	\left(K_b^{a}\right)_{u,v} &= 2\tau_i\rho_{x^\ast}\delta_{a,b} + Q_{1}^{a}\! -\! \frac{1}{d}\Tr\left[Q_{1}^{a}\right]\id - \sum_{u,v} \left(\Gamma_{b}^{a}\right)_{u,v} \quad \text{if} \ u+v=1 \\
	\left(K_b^{a}\right)_{u,v} &= \tau_i\rho_{x^\ast}\!\left[\left(1\!-\!t_i\right)\delta_{a,b}\!+\!t_i\right] + Q_{2}^{a}\! -\! \frac{1}{d}\Tr\left[Q_{2}^{a}\right]\id - \sum_{u,v} \left(\Gamma_{b}^{a}\right)_{u,v} \quad \text{if} \ u+v=2 \\
	\left(K_b^{a}\right)_{u,v} &= Q_{\omega}^{a}\! -\! \frac{1}{d}\Tr\left[Q_{\omega}^{a}\right]\id - \sum_{u,v} \left(\Gamma_{b}^{a}\right)_{u,v} \quad \text{if} \ u+v=\omega \ \text{for} \ \omega > 2
\end{aligned}
\end{equation}
collecting all terms in $\mathcal{L}_i$ multiplying the primal variables. The infimum $\mathcal{I}$ represents a lower bound on the optimal solution of the primal semidefinite program. However, note that the infimum will diverge unless $K_b^{a}=0$. To get good bounds therefore, we must maximise $\mathcal{I}$ over the set of dual variables that satisfy $K_b^{a}=0$. This maximisation leads to the dual formulation of the semidefinite program,
\begin{equation}\label{eq:shannon_dual}
\begin{aligned}
    c_m + \sum_{i=1}^{m-1} \ \underset{\Gamma^{a}_{b}, s, R, Q_{\omega}^{a}}{\text{maximize}} & \quad \left(- sW - \Tr\left[R\right]\right)  \\
\text{subject to} & \quad \Gamma_{b}^{a} \succeq 0 \\
& \quad \sum_{a} \left(\Gamma_{b}^{a}\right)_{0,0} = s\sum_{x}\frac{1}{16}\delta_{b,x}\rho_x + R  \\
& \quad \sum_{u,v} \left(\Gamma_{b}^{a}\right)_{u,v} = 2\tau_i\rho_{x^\ast}\delta_{a,b} + Q_{1}^{a}\! -\! \frac{1}{d}\Tr\left[Q_{1}^{a}\right]\id \quad \text{if} \ u+v=1 \\
& \quad \sum_{u,v} \left(\Gamma_{b}^{a}\right)_{u,v} = \tau_i\rho_{x^\ast}\!\left[\left(1\!-\!t_i\right)\delta_{a,b}\!+\!t_i\right] + Q_{2}^{a}\! -\! \frac{1}{d}\Tr\left[Q_{2}^{a}\right]\id \quad \text{if} \ u+v=2 \\
& \quad \sum_{u,v} \left(\Gamma_{b}^{a}\right)_{u,v} = Q_{\omega}^{a}\! -\! \frac{1}{d}\Tr\left[Q_{\omega}^{a}\right]\id \quad \text{if} \ u+v=\omega \ \text{for} \ \omega > 2 \ .
\end{aligned}
\end{equation}
Any set of $kd\times kd$ (for $2k$ being the order of the semidefinite program relaxation) positive semidefinite matrices $\Gamma_{b}^{a}$, $d\times d$ matrices $R$ and $Q_{\omega}^{a}$ and scalars $s$ that satisfy the dual constraints above, represent a valid lower bound on the Shannon entropy. \\

\textbf{Results.} The prepare-and-measure experiment consists of a total of $17$ distinct state preparations, $16$ $4-$dimensional SIC states plus a maximally mixed state from which we wish to extract randomness, i.e.~$\rho_{x^\ast}=\frac{\id}{4}$. The experiment runs in a lab over $N=1196436$ rounds and the observed frequencies are stored, from which we compute observable state discrimination success probability $W_{\text{obs}}=0.24730$.  We compute the min-entropy bounding the guessing probability through the semidefinite program in \eqref{eq:pgsdp_primal} and obtain $H_{min}=2.740$. We then compute the Shannon entropy with the primal and dual semidefinite programs in \eqref{eq:shannon_primal} and \eqref{eq:shannon_dual} respectively. In Tab.~\ref{tab:shannon} we show the certifiable Shannon entropies with different Gauss-Radau quadratures ($m$) and orders of relaxation ($2k$).

To bound the certifiable randomness, we use the entropy accumulation theorem \cite{metger2022} (see Ref.~\cite{Carceller25} for a similar approach). Concretely, this states that the amount of certifiable randomness per-round, quantified through the \textit{smooth min-entropy} $H^{\varepsilon}_{\min}$ (i.e.~the maximum min-entropy for any state $\varepsilon$-close to a fixed state \cite{renner2006}), is bounded by the following quantity
\begin{align}
\label{eq:GEAT}
    \frac{1}{N}H_{\min}^{\varepsilon}(B^{N}|E^{N}) \geq \underset{c_{N}\in\Omega}{\min} f_\text{min}\left(\text{freq.}(c_{N})\right) - \frac{\alpha-1}{2-\alpha}\frac{\text{ln}(2)}{2}V^2 - \frac{1}{N}\frac{g(\varepsilon)+\alpha\log\left(1/\text{Pr}\left[\Omega\right]\right)}{\alpha-1} - \left(\frac{\alpha-1}{2-\alpha}\right)^{2}K'(\alpha) \ .
\end{align}
Here, $f_\text{min}$ is a so-called \textit{min-tradeoff function} and it is defined as an analytical function on the observed frequencies (freq.$(c_N)$) computed on the events whenever randomness is certified ($c_{N}$) such that it lower-bounds the minimum von Neumann (or, in our case, Shannon) entropy. Namely,
\begin{align}
\label{eq:trade_off}
	f_\text{min}\left(q\right)  \leq \underset{\nu\in\Sigma_i(q)}{\min} H(B_i|E_i)_\nu
\end{align}
for $\Sigma_i(q)$ being the set of states that can be generated after the measurement at the $i^{th}$ experiment round given the observable frequencies $q$. The lower bound in \eqref{eq:GEAT} is computed on the minimum $f_\text{min}$ over the total number of observed events $c_{N}$ belonging to a particular chosen event $\Omega$ which, in our case, corresponds to the events where we certify randomness, i.e. whenever the maximally mixed state is prepared, occurring with probability Pr$(\Omega)=1/5$. Here, we take as a min-tradeoff function the optimal objective function of the dual semidefinite program in \eqref{eq:shannon_dual}. That is,
\begin{align}
f_\text{min}(W) =  c_m + \tilde{r} + \tilde{s} \ W 
\quad \text{for} \quad \tilde{s} = -\sum_{i=1}^{m-1}s_i^\ast  , \ \tilde{r} = -\sum_{i=1}^{m-1}\Tr\left[R_i^\ast\right] \ ,
\end{align}
where the observable frequencies are concentrated into the observable witness $W$, and $s_i^\ast$ and $R_i^\ast$ represent feasible points of the dual semidefinite program in \eqref{eq:shannon_dual} for each iteration $i$. For any observable linear witness $W$, the function $f_\text{min}(W)$ is always a lower-bound on the minimum Shannon entropy. Moreover, from the min-tradeoff function we need to compute the maximum $\text{Max}(f_\text{min})=f_\text{min}(W_{\text{obs}}+\text{Var}(W_{\text{obs}}))$, minimum $\text{Min}(f_\text{min})=f_\text{min}(W_{\text{obs}}-\text{Var}(W_{\text{obs}}))$ and its variance $\text{Var}(f_\text{min})=\tilde{s}^2 \text{Var}(W_{\text{obs}})$, with $W_{\text{obs}}$ being the observed witness, and $\text{Var}(W_{\text{obs}})=5\cdot 10^{-5}$ its variance. These are then used to compute the following quantities
\begin{align}
g(\varepsilon) =& -\log(1-\sqrt{1-\varepsilon^2}) \ , \\
V =& \log(2 d_A^2 + 1) + \sqrt{2+\text{Var}(f_\text{min})} \ , \\
K'(\alpha) =& \frac{(2-\alpha)^3}{6(3-2\alpha)^3\text{ln}(2)}2^{\frac{\alpha-1}{2-\alpha}(2\log d_A+\text{Max}(f_\text{min})-\text{Min}(f_\text{min}))}\text{ln}^{3}\left(2^{2\log d_A+\text{Max}(f_\text{min})-\text{Min}(f_\text{min})}+\rme^2\right) \ ,
\end{align}
with $d_A=4$ corresponding to the dimension of the SIC state preparations. Finally, the bound in \eqref{eq:GEAT} is originally derived using some appropiate properties of the Reny entropies $H_{\alpha}$ for $\alpha\in\left(1,2\right)$. We thus set $\alpha=1+1/\sqrt{N}$ which is entirely motivated by Ref~\cite{metger2022}, so that one gets a correction term $\mathcal{O}(1/\sqrt{N})$ in the bound on the certifiable entropy per-round.

After running the dual semidefinite program from \eqref{eq:shannon_dual} with a quadrature level $m=18$ and a relaxation up to order $2k=10$, we obtain a min-tradeoff function characterized by the quadrature set value $c_m=9.2305$, and optimal parameters $\tilde{r}=-65.2748$ and $\tilde{s}=238.8474$, obtaining a min-tradeoff function evaluated with $W_{\text{obs}}=0.24730$ of $f_\text{min}(W_{\text{obs}})=3.0227$. As a result therefore, with $\varepsilon=10^{-4}$, we obtain a correction of $-0.0441$ which amounts to a total of $2.9786$ bits of randomness per round.

\bibliography{all_references}